\documentclass[12pt,a4paper]{article}
\pdfoutput=1
\usepackage{graphicx}
\usepackage[T1]{fontenc}
\usepackage[utf8]{inputenc}
\usepackage{textcomp}
\usepackage[sc,osf]{mathpazo}
\usepackage{a4wide}  
\usepackage{latexsym,amsthm,amsfonts,amsmath,mathrsfs,amssymb}
\usepackage{booktabs} % for much better looking tables
\usepackage[unicode,implicit]{hyperref}
\hypersetup{%
  pdftitle    = {Non-Abelian Rotating Black Holes in 4- and 5-Dimensional
    Gauged Supergravity}
  pdfkeywords = {gravity, supergravity, gauge, non-Abelian, black hole, rotating, extremal, entropy, instanton, dyon}
  pdfauthor   = {Tomas Ortin, Alejandro Ruiperez},
  plainpages  = true,
  colorlinks  = true,
  citecolor   = blue,
  urlcolor    = red, 
  linkcolor   = black
}
\newcommand{\hepth}[1]{{\tt
hep-th/\href{http://www.arXiv.org/abs/hep-th/#1}{#1}}}

\newcommand{\arxiv}[1]{{\tt arXiv:\href{http://www.arXiv.org/abs/#1}{#1}}}
\newcommand{\doi}[1]{{\tt DOI:\href{http://dx.doi.org/#1}{#1}}}

\makeatletter
\@addtoreset{equation}{section}
\makeatother

\begin{document}

\begin{flushright}
\small
IFT-UAM/CSIC-19-14\\
%\texttt{arXiv:yymm.nnnnn [hep-th]}\\
April 30\textsuperscript{th}, 2019\\
\normalsize
\end{flushright}

\vspace{1.5cm}

\begin{center}

{\Large {\bf Non-Abelian Rotating Black Holes\\[.5cm] in 4- and 5-Dimensional
    Gauged Supergravity}}

\vspace{1.5cm}

\renewcommand{\thefootnote}{\alph{footnote}}
{\sl\large Tom\'{a}s Ort\'{\i}n$^{~1,}$\footnote{Email: {\tt tomas.ortin[at]csic.es}}
 and Alejandro Ruip\'erez$^{~1,2,}$}\footnote{Email: {\tt alejandro.ruiperez[at]uam.es}}

\setcounter{footnote}{0}
\renewcommand{\thefootnote}{\arabic{footnote}}

\vspace{0.5cm}

${}^{1}${\it Instituto de F\'{\i}sica Te\'orica UAM/CSIC,\\
C/ Nicol\'as Cabrera, 13--15,  C.U.~Cantoblanco, E-28049 Madrid, Spain} 

\vspace{0.2cm}

${}^{2}${\it Instituut voor Theoretische Fysica,\\
  KU Leuven, Celestijnenlaan 200D, B-3001 Leuven, Belgium}

\vspace{1.8cm}

%%%%%%%%%%%%%%%%%%%%%%%%%%%%%%%%%%%%%%%%%%%%%%%%%%%%%%%%%%%%%%%%%%%%%%

{\bf Abstract}
\end{center}
\begin{quotation}
  {\small We present new supersymmetric black-hole solutions of the 4- and 5-dimensional
    gauged supergravity theories that one obtains by dimensional reduction on
    $T^{5}$ and $T^{6}$ of Heterotic supergravity with a triplet of Yang-Mills
    fields. The new ingredient of our solutions is the presence of dyonic
    non-Abelian fields which allows us to obtain a generalization of the BMPV
    black hole with two independent angular momenta and the first example of a
   supersymmetric, rotating, asymptotically-flat black hole with a regular horizon 
   in 4 dimensions.}
\end{quotation}

\newpage
%%%%%%%%%%%%%%%%%%%%%%%%%%%%%%%%%%%%%%%%%%%%%%%%%%%%%%%%%%%%%%%%%%%%%%
%%%%%%%%%%%%%%%%%%%%%%%%%%%%%%%%%%%%%%%%%%%%%%%%%%%%%%%%%%%%%%%%%%%%%%
%%%%%%%%%%%%%%%%%%%%%%%%%%%%%%%%%%%%%%%%%%%%%%%%%%%%%%%%%%%%%%%%%%%%%%
%%%%%%%%%%%%%%%%%%%%%%%%%%%%%%%%%%%%%%%%%%%%%%%%%%%%%%%%%%%%%%%%%%%%%%
\pagestyle{plain}
%%%%%%%%%%%%%%%%%%%%%%%%%%%%%%%%%%%%%%%%%%%%%%%%%%%%%%%%%%%%%%%%%%%%%%
%%%%%%%%%%%%%%%%%%%%%%%%%%%%%%%%%%%%%%%%%%%%%%%%%%%%%%%%%%%%%%%%%%%%%%
%%%%%%%%%%%%%%%%%%%%%%%%%%%%%%%%%%%%%%%%%%%%%%%%%%%%%%%%%%%%%%%%%%%%%%
%%%%%%%%%%%%%%%%%%%%%%%%%%%%%%%%%%%%%%%%%%%%%%%%%%%%%%%%%%%%%%%%%%%%%%

\tableofcontents

%\newpage

%%%%%%%%%%%%%%%%%%%%%%%%%%%%%%%%%%%%%%%%%%%%%%%%%%%%%%%%%%%%%%%%%%%%%%
%%%%%%%%%%%%%%%%%%%%%%%%%%%%%%%%%%%%%%%%%%%%%%%%%%%%%%%%%%%%%%%%%%%%%%
%%%%%%%%%%%%%%%%%%%%%%%%%%%%%%%%%%%%%%%%%%%%%%%%%%%%%%%%%%%%%%%%%%%%%%
%%%%%%%%%%%%%%%%%%%%%%%%%%%%%%%%%%%%%%%%%%%%%%%%%%%%%%%%%%%%%%%%%%%%%%
\section{Introduction}
%%%%%%%%%%%%%%%%%%%%%%%%%%%%%%%%%%%%%%%%%%%%%%%%%%%%%%%%%%%%%%%%%%%%%%
%%%%%%%%%%%%%%%%%%%%%%%%%%%%%%%%%%%%%%%%%%%%%%%%%%%%%%%%%%%%%%%%%%%%%%
%%%%%%%%%%%%%%%%%%%%%%%%%%%%%%%%%%%%%%%%%%%%%%%%%%%%%%%%%%%%%%%%%%%%%%
%%%%%%%%%%%%%%%%%%%%%%%%%%%%%%%%%%%%%%%%%%%%%%%%%%%%%%%%%%%%%%%%%%%%%%

The study of the classical solutions of General Relativity and its
generalizations has been one of the major sources of information about the
properties of those theories. Supergravity theories are a particularly
interesting class of generalizations of General Relativity because many of
them describe low-energy effective field theories of different superstring
theories and a great deal of work has been devoted to them and their classical
solutions (specially to the supersymmetric ones and specially to those
describing black holes).\footnote{For a review of superstring theories and
  their classical solutions from this point of view see
  Ref.~\cite{Ortin:2015hya}.}

One of the most important features of these theories is the presence of vector
and scalar fields that give rise to many interesting phenomena and properties
of the black-holes solutions. The electric and magnetic charges associated to
those vectors play a crucial role in the stringy interpretation of the black
holes that carry them and determine completely their entropy formula in the
static, extremal cases. However, in most of the literature, only models with
Abelian vector fields have been considered, even though non-Abelian vector
fields play a most relevant role in our current understanding of Nature and
most string models (specially the more realistic ones) include them in their
spectra. In the case of the Heterotic Superstring, non-Abelian vector fields occur at
first order in $\alpha'$ and can play an important role in the suppression or
even cancellation of $\alpha'$ corrections to black-hole geometries
\cite{Cano:2018qev,Chimento:2018kop,Cano:2018brq}. Thus, it is clearly
important to study the interplay between gravity and Yang-Mills fields in this
context and to understand how the results obtained in the Abelian case are
modified by the presence of the later.

During the last decade our group has been trying to fill this gap in our
understanding exploring supersymmetric solutions (specially on black-hole or
black-ring solutions) with non-Abelian fields in gauged supergravity theories
\cite{Cariglia:2004kk,Huebscher:2007hj,Bellorin:2007yp,Meessen:2008kb,Hubscher:2008yz,Bellorin:2008we,Meessen:2012sr,Bueno:2014mea,Meessen:2015nla,Meessen:2015enl,Ortin:2016sih,Cano:2016rls,Chimento:2017tsv,Meessen:2017rwm,Chimento:2018elo,Cano:2017sqy,Ortin:2016bnl}.
They have also played an important role in the construction of black-hole
microstate geometries \cite{Ramirez:2016tqc,Avila:2017pwi}\footnote{For a
  review on black-hole and black-ring microstate geometries see, for instance,
  Ref.~\cite{Bena:2007kg} and references therein.} and, furthermore, they have
turned out to be the intermediate step necessary to construct
$\alpha'$-corrected stringy supersymmetric black-hole solutions
\cite{Cano:2017qrq,Cano:2018qev,Chimento:2018kop,Cano:2018brq,Cano:2018hut}.
This is due to the fact that, in the context of the Heterotic Superstring
effective action, the non-Abelian gauge fields occur at first order in
$\alpha'$ and it is known that the curvature squared of the torsionful spin
connection occurs at the same order just as another gauged field
\cite{Bergshoeff:1988nn}, a feature of the theory that makes the Green-Schwarz
anomaly-cancellation mechanism possible \cite{Green:1984sg}. The same
mechanism can be used to cancel some of the $\alpha'$ corrections as well.

Although the multicenter solutions constructed in Ref.~\cite{Meessen:2017rwm}
have angular momentum, the only rotating, supersymmetric, single-center
solutions constructed so far with non-Abelian fields are the black rings and
black holes of Ref.~\cite{Ortin:2016bnl} which are the simplest
generalizations of the Abelian ones. In particular, the rotating black-hole
solution presented in that reference can be understood as a BMPV black hole
(only one independent angular momentum) with additional non-Abelian hair.

The aim of this work is to extend the catalogue of known rotating,
single-center, black-hole solutions in 4 and 5 dimensions by
exploring the effect of adding dyonic non-Abelian fields
defined on hyperK\"ahler spaces. The supergravity theories that we are
going to consider in this paper are the 4- and 5-dimensional versions of the
SU$(2)$-gauged $\rm{ST[2,6]}$ model (an extension of the STU model),
whose main interest lies in the fact that it can be obtained by toroidal
compactification and truncation of the $\mathcal{N}=1,d=10$
supergravity coupled to non-Abelian vector fields \cite{Cano:2017qrq}, 
which is often referred to as Heterotic supergravity. 

The dyonic non-Abelian fields that we will consider are
generalizations of those presented in Refs.~\cite{Ramirez:2016tqc} and
\cite{Lambert:1999ua}. The latter were used in Ref.~\cite{Eyras:2000dg} to construct a
globally smooth solution of Heterotic supergravity.
The solutions that we are going to consider here generalize that one
by considering non-trivial hyperK\"ahler base spaces on which the
dyonic instanton is defined and by turning on additional fields that
give rise to regular event horizons, which in 4 dimensions
violates the no-go theorem for regular, supersymmetric, rotating black
holes proven in Ref.~\cite{Bellorin:2006xr}. In 5 dimensions, they allow us to find
supersymmetric black holes with two independent angular momenta.

This paper is organized as follows. In Section~\ref{sec-setup} we
describe the class of gauged $\mathcal{N}=1,d=5$ theories (8
supercharges) we are going to work with and a general solution-generating
technique for supersymmetric solutions of this kind of theories. 
In Section~\ref{sec-gensolST[2,6]} we apply this technique to the particular
model we are interested in, the ST$[2,6]$ model, and discuss which kind of dyonic
non-Abelian fields, in particular, can be added to it. We also
describe the dimensional reduction of this model (and of the
corresponding solutions) to 4 dimensions, where the theory becomes
a model of gauged $\mathcal{N}=2,d=4$ supergravity. In
Section~\ref{explicit-solutions} we focus on the study of single-center
black holes both in 5 and 4 dimensions. 
Section~\ref{sec-discussion} contains a discussion of our results.

%%%%%%%%%%%%%%%%%%%%%%%%%%%%%%%%%%%%%%%%%%%%%%%%%%%%%%%%%%%%%%%%%%%%%%
%%%%%%%%%%%%%%%%%%%%%%%%%%%%%%%%%%%%%%%%%%%%%%%%%%%%%%%%%%%%%%%%%%%%%%
%%%%%%%%%%%%%%%%%%%%%%%%%%%%%%%%%%%%%%%%%%%%%%%%%%%%%%%%%%%%%%%%%%%%%%
%%%%%%%%%%%%%%%%%%%%%%%%%%%%%%%%%%%%%%%%%%%%%%%%%%%%%%%%%%%%%%%%%%%%%%
\section{The general set up}
\label{sec-setup}
%%%%%%%%%%%%%%%%%%%%%%%%%%%%%%%%%%%%%%%%%%%%%%%%%%%%%%%%%%%%%%%%%%%%%% 
%%%%%%%%%%%%%%%%%%%%%%%%%%%%%%%%%%%%%%%%%%%%%%%%%%%%%%%%%%%%%%%%%%%%%%
%%%%%%%%%%%%%%%%%%%%%%%%%%%%%%%%%%%%%%%%%%%%%%%%%%%%%%%%%%%%%%%%%%%%%%
%%%%%%%%%%%%%%%%%%%%%%%%%%%%%%%%%%%%%%%%%%%%%%%%%%%%%%%%%%%%%%%%%%%%%%

%%%%%%%%%%%%%%%%%%%%%%%%%%%%%%%%%%%%%%%%%%%%%%%%%%%%%%%%%%%%%%%%%%%%%%
%%%%%%%%%%%%%%%%%%%%%%%%%%%%%%%%%%%%%%%%%%%%%%%%%%%%%%%%%%%%%%%%%%%%%%
%%%%%%%%%%%%%%%%%%%%%%%%%%%%%%%%%%%%%%%%%%%%%%%%%%%%%%%%%%%%%%%%%%%%%%
%%%%%%%%%%%%%%%%%%%%%%%%%%%%%%%%%%%%%%%%%%%%%%%%%%%%%%%%%%%%%%%%%%%%%%
\subsection{$\mathcal{N}=1,d=5$ Super-Einstein-Yang-Mills theories}
%%%%%%%%%%%%%%%%%%%%%%%%%%%%%%%%%%%%%%%%%%%%%%%%%%%%%%%%%%%%%%%%%%%%%%
%%%%%%%%%%%%%%%%%%%%%%%%%%%%%%%%%%%%%%%%%%%%%%%%%%%%%%%%%%%%%%%%%%%%%%
%%%%%%%%%%%%%%%%%%%%%%%%%%%%%%%%%%%%%%%%%%%%%%%%%%%%%%%%%%%%%%%%%%%%%%
%%%%%%%%%%%%%%%%%%%%%%%%%%%%%%%%%%%%%%%%%%%%%%%%%%%%%%%%%%%%%%%%%%%%%%

$\mathcal{N}=1,d=5$ super-Einstein-Yang-Mills (SEYM) theories are the
simplest theories of 5-dimensional supergravity containing non-Abelian
gauge fields.  They can be described as the simplest and minimal
supersymmetrization of 5-dimensional Einstein-Yang-Mills theories or
as the simplest coupling of 5-dimensional super-Yang-Mills theory to
supergravity.

For our purposes, it is convenient to describe these theories as the
result of gauging a subgroup of the isometry group of the scalar
manifold of a $\mathcal{N}=1,d=5$ supergravity coupled to vector
multiplets.\footnote{Our conventions are those of
  Refs.~\cite{Bellorin:2006yr,Bellorin:2007yp}, which are based on
  Ref.~\cite{Bergshoeff:2004kh}. }  Therefore, these theories describe

\begin{enumerate}
\item The supergravity multiplet containing the graviton $e^{a}{}_{\mu},$ the
gravitino $\psi_{\mu}^{i}$ and the graviphoton $A^{0}{}_{\mu}$  
\item $n_{v}$ vector multiplets labeled by $x=1,\cdots,n_{v}$ (each
  containing a real vector field $A^{x}{}_{\mu},$ a real scalar
  $\phi^{x}$ and a gaugino $\lambda^{i\, x}$).
\end{enumerate}

The above field content does not determine completely the theory,
since the matter fields can be coupled to supergravity in different
ways, even before gauging. In order to describe the different
possibilities, it is convenient to combine the indices of the matter
vector fields and of the graviphoton into a single index 
$I,J,\ldots=0,1,\cdots,n_{v}$ so all the vector fields are denoted by
a single object $A^{I}{}_{\mu}$. Then, all the couplings between the
fields of a given ungauged theory (between scalars, ${\mathfrak g}_{xy}(\phi)$,
between scalars and vectors $a_{IJ}(\phi)$ and the Chern-Simons
couplings between vectors) are completely determined by a constant,
completely symmetric tensor $C_{IJK}$.\footnote{Our description of
  these theories will be minimal, giving only the information required
  to obtain and explain the results presented in this paper. The
  reader interested in further details on these theories, such as how
  to derive the couplings between the fields from $C_{IJK}$, may
  consult the references quoted above.}

Generically, an ungauged theory of $\mathcal{N}=1,d=5$ supergravity
coupled to vector multiplets will be invariant under certain group of
symmetries acting only on the vector and scalar fields.\footnote{Here
  we are ignoring R-symmetry.} The action of these symmetries on the
scalars has to preserve $g_{xy}(\phi)$, the metric of the scalar
manifold and, therefore, it will act on them as the isometries
generated by the Killing vectors, that we will label by
$k_{I}{}^{x}(\phi)$, and which can vanish for some values of $I$. At
the same time, because of the non-trivial couplings between scalar and
vector fields, the vectors will be rotated by some given matrices.

In many cases, it is possible to gauge a (necessarily non-Abelian)
subgroup of this symmetry group using as gauge fields a subset of the
vector fields of the theory. We will denote the structure constants of
the gauge group by $f_{IJ}{}^{K}$ using the convention that they and
the associated Killing vectors, will just vanish for the values of the
indices that do not correspond to the gauge fields. In the gauging
procedure, the partial derivatives of the scalars are promoted to
gauge-covariant derivatives
$\mathfrak{D}_{\mu}\phi^{x}=
\partial_{\mu}\phi^{x}+gA^{I}{}_{\mu}k_{I}{}^{x}$ and the Abelian
vector field strengths are promoted to their non-Abelian counterparts
$F^{I}{}_{\mu\nu}=2\partial_{[\mu}A^{I}{}_{\nu]}
+gf_{JK}{}^{I}A^{J}{}_{\mu}A^{K}{}_{\nu}$. Here $g$ is the gauge
coupling constant. Gauge symmetry also demands the addition of further
terms to the Chern-Simons terms, but, as different from what happens
in the gauging of many other supergravity theories, supersymmetry does
not demand the addition of a scalar potential and no effective
cosmological constant is present in the theory and its solutions.

The bosonic action of these gauged supergravities, that we call
$\mathcal{N}=1,d=5$ Super-Einstein-Yang-Mills (SEYM) theories, is given by 

\begin{equation}
  \label{eq:genericN1d5SEYMaction}
\begin{array}{rcl}
  S
  & = &
        {\displaystyle
        \frac{1}{16\pi G_{N}^{(5)}}\int
        } d^{5}x\sqrt{|g|}\
\biggl\{
R
+{\textstyle\frac{1}{2}}{\mathfrak g}_{xy}\mathfrak{D}_{\mu}\phi^{x}
\mathfrak{D}^{\mu}\phi^{y}
-{\textstyle\frac{1}{4}} a_{IJ} F^{I\, \mu\nu}F^{J}{}_{\mu\nu}
\\ \\ & & 
+\tfrac{1}{12\sqrt{3}}C_{IJK}
{\displaystyle\frac{\varepsilon^{\mu\nu\rho\sigma\alpha}}{\sqrt{|g|}}}
\left[
F^{I}{}_{\mu\nu}F^{J}{}_{\rho\sigma}A^{K}{}_{\alpha}
-\tfrac{1}{2}gf_{LM}{}^{I} F^{J}{}_{\mu\nu} 
A^{K}{}_{\rho} A^{L}{}_{\sigma} A^{M}{}_{\alpha}
\right.
  \\ \\ & & 
\left.
+\tfrac{1}{10} g^{2} f_{LM}{}^{I} f_{NP}{}^{J} 
A^{K}{}_{\mu} A^{L}{}_{\nu} A^{M}{}_{\rho} A^{N}{}_{\sigma} A^{P}{}_{\alpha}
\right]
\biggr\}\, ,
\end{array}
\end{equation}

\noindent
where $G_{N}^{(5)}$ is the 5-dimensional Newton constant and $g$ is the Yang-Mills coupling constant.\footnote{The symbol $|g|$ denotes, however, the determinant of the 5-dimensional metric $g_{\mu\nu}=e^a{}_{\mu}e^b{}_{\nu}\eta_{ab}$.}

For the sake of completeness and also for their use in defining the
charges of the solutions, we quote the equations of motion that follow
from this action:

\begin{eqnarray}
\mathcal{E}_{\mu\nu} 
& \equiv &
\frac{1}{2\sqrt{g}} e_{a(\mu}
\frac{\delta S}{\delta e_{a}{}^{\nu)}}
\nonumber \\
& & \nonumber \\
& = &
G_{\mu\nu}
-{\textstyle\frac{1}{2}}a_{IJ}\left(F^{I}{}_{\mu}{}^{\rho} F^{J}{}_{\nu\rho}
-{\textstyle\frac{1}{4}}g_{\mu\nu}F^{I\, \rho\sigma}F^{J}{}_{\rho\sigma}
\right)      
\nonumber\\
& & \nonumber  \\
& & 
+{\textstyle\frac{1}{2}}{\mathfrak g}_{xy}\left(\mathfrak{D}_{\mu}\phi^{x} 
\mathfrak{D}_{\nu}\phi^{y}
-{\textstyle\frac{1}{2}}g_{\mu\nu}
\mathfrak{D}_\rho\phi^{x} \mathfrak{D}^{\rho}\phi^{y}\right)
\\ 
  & & \nonumber \\
  \label{eq:vectoreom}
\mathcal{E}_{I}{}^{\mu} 
& \equiv &
\frac{1}{\sqrt{g}}\frac{\delta S}{\delta A^{I}{}_{\mu}}
\nonumber \\
& & \nonumber \\
& = & 
\mathfrak{D}_{\nu}\left(a_{IJ} F^{J\, \nu\mu}\right)
+{\textstyle\frac{1}{4\sqrt{3}}} 
\frac{\varepsilon^{\mu\nu\rho\sigma\alpha}}{\sqrt{g}}
C_{IJK} F^{J}{}_{\nu\rho}F^{k}{}_{\sigma\alpha}
+g k_{I\,x} \mathfrak{D}^{\mu}\phi^{x}
\\
& & \nonumber \\
\mathcal{E}^{x} 
& \equiv &
-\frac{{\mathfrak g}^{xy}}{\sqrt{g}}
\frac{\delta S}{\delta \phi^{y}}
\nonumber \\
& & \nonumber \\
& = & 
\mathfrak{D}_{\mu}\mathfrak{D}^{\mu}\phi^{x} 
+{\textstyle\frac{1}{4}}{\mathfrak g}^{xy} \partial_{y}
a_{IJ} F^{I\, \rho\sigma}F^{J}{}_{\rho\sigma}.
\end{eqnarray}

%%%%%%%%%%%%%%%%%%%%%%%%%%%%%%%%%%%%%%%%%%%%%%%%%%%%%%%%%%%%%%%%%%%%%%
%%%%%%%%%%%%%%%%%%%%%%%%%%%%%%%%%%%%%%%%%%%%%%%%%%%%%%%%%%%%%%%%%%%%%%
%%%%%%%%%%%%%%%%%%%%%%%%%%%%%%%%%%%%%%%%%%%%%%%%%%%%%%%%%%%%%%%%%%%%%%
%%%%%%%%%%%%%%%%%%%%%%%%%%%%%%%%%%%%%%%%%%%%%%%%%%%%%%%%%%%%%%%%%%%%%%
\subsection{A solution-generating technique for $\mathcal{N}=1, d=5$
  SEYM theories}
%%%%%%%%%%%%%%%%%%%%%%%%%%%%%%%%%%%%%%%%%%%%%%%%%%%%%%%%%%%%%%%%%%%%%%
%%%%%%%%%%%%%%%%%%%%%%%%%%%%%%%%%%%%%%%%%%%%%%%%%%%%%%%%%%%%%%%%%%%%%%
%%%%%%%%%%%%%%%%%%%%%%%%%%%%%%%%%%%%%%%%%%%%%%%%%%%%%%%%%%%%%%%%%%%%%%
%%%%%%%%%%%%%%%%%%%%%%%%%%%%%%%%%%%%%%%%%%%%%%%%%%%%%%%%%%%%%%%%%%%%%%

Classical solutions of $\mathcal{N}=1, d=5$ SEYM theories with a
symmetric scalar manifold can be constructed using the following
building blocks:\footnote{This recipe stems from the characterization
  of timelike supersymmetric solutions of the most general
  $\mathcal{N}=1, d=5$ supergravity theory including vector
  supermultiplets and hypermultiplets and generic gaugings made in
  Ref.~\cite{Bellorin:2007yp}, based in the results of
  Ref.~\cite{Bellorin:2006yr}. The inclusion of tensor supermultiplets
  was considered in Ref.~\cite{Bellorin:2008we}. We have adapted those
  results to the case at hands. Furthermore, we have restricted this
  recipe to models with a symmetric scalar manifold, for simplicity
  (the model we are going to study belongs to this class). In these
  models, but not in general, the tensor $C^{IJK}$ that one obtains by
  raising the indices with the inverse of $a_{IJ}(\phi)$
  ($a^{IJ}(\phi)$) is constant and identical to $C_{IJK}$.}

\begin{enumerate}

\item A 4-dimensional hyperK\"ahler (HK) manifold with metric
  $d\sigma^{2}=h_{\underline{m}\underline{n}}\, dx^{m} dx^{n}$\footnote{In
  our conventions $m,n=1,\dots, 4$ are tangent space indices whereas 
  $\underline{m},\underline{n}=1,\dots, 4$ are curved indices.} 
 following fields defined on it:

\item $n_{v}+1$ vector fields
  $\hat{A}^{I}=\hat{A}^{I}_{\underline{m}}\,dx^{m}$.

\item $n_{v}+1$ functions $Z_{I}$, $I=0,\ldots, n_{v}$ defined

\item A 1-form $\omega=\omega_{\underline {m}} \,dx^{m}$.

\end{enumerate}

In terms of these building blocks and the tensor $C_{IJK}$ that
defines the model, the 5-dimensional physical fields (metric
$g_{\mu\nu}$, vector fields $A^{I}$ and scalar fields $\phi^{x}$) are
given by

\begin{eqnarray}
  \label{eq:gen-metric}
  ds^{2}
  & = &
        f^{2}\left(dt+\omega\right)^{2}-f^{-1}d\sigma^{2} \, ,
  \\
  \nonumber\\
  \label{eq:gen-vectors}
  A^{I}
  & = &
        -27\sqrt{3}\, C^{IJK}Z_{J} Z_{K} f^{3} \left(dt+\omega\right)
        +\hat{A}^{I}\, ,
  \\
  \nonumber\\
  \label{eq:scalarsp1}
  \phi^{x}
  & = &
        \frac{Z_{x}}{Z_{0}}\, ,
\end{eqnarray}

\noindent
where the metric function $f$ is given by

\begin{equation}
f^{-3}=27 C^{IJK}Z_{I} Z_{J} Z_{K}\, .
\end{equation}

The building blocks of the solution ($\hat{A}^{I}, Z_{I}, \omega$)
must satisfy the following differential equations on the
HK manifold:

\begin{eqnarray}
\label{eq:selfduality}
  \hat{F}^{I}
  & = &
        \star_{\sigma} \hat{F}^{I}\, ,
  \\
\nonumber\\
\label{eq:Zs}
  \hat{\mathfrak{D}}\star_{\sigma} \hat{\mathfrak{D}} Z_{I}
  & = &
        -\tfrac{1}{3}C_{IJK}\,\hat{F}^{I}\wedge\hat{F}^{J}\, ,
  \\
\nonumber\\
\label{eq:omega}
  d\omega+\star_{\sigma} d\omega
  & = &
        \sqrt{3} Z_{I} \hat{F}^{I}\, ,
\end{eqnarray}

\noindent
where $\star_{\sigma}$ is the restriction of the Hodge star to the
4-dimensional HK metric $d\sigma^{2}$, $\hat{\mathfrak{D}}$
is the gauge-covariant derivative with respect to the hatted gauge
connection $\hat{A}^{I}$

\begin{equation}
\hat{\mathfrak{D}} Z_{I}=d Z_{I} +g f_{IJ}{}^{K} \hat{A}^{J}\wedge Z_{K}\, ,
\end{equation}

\noindent
and $\hat{F}^{I}$ is the field strength of that connection

\begin{equation}
  \hat{F}^{I}
  =
  d \hat{A}^{I}+\frac{g}{2} f_{JK}{}^{I} \hat{A}^{J} \wedge \hat{A}^{K}\, .
\end{equation}

The solutions constructed in this way are time-independent and, in
general ($\omega\neq 0$) stationary. They are also (``timelike'')
supersymmetric and preserve $1/2$ of the 8 supersymmetries of these
theories.

%%%%%%%%%%%%%%%%%%%%%%%%%%%%%%%%%%%%%%%%%%%%%%%%%%%%%%%%%%%%%%%%%%%%%%
%%%%%%%%%%%%%%%%%%%%%%%%%%%%%%%%%%%%%%%%%%%%%%%%%%%%%%%%%%%%%%%%%%%%%%
%%%%%%%%%%%%%%%%%%%%%%%%%%%%%%%%%%%%%%%%%%%%%%%%%%%%%%%%%%%%%%%%%%%%%%
%%%%%%%%%%%%%%%%%%%%%%%%%%%%%%%%%%%%%%%%%%%%%%%%%%%%%%%%%%%%%%%%%%%%%%
\section{Dyonic solutions of the $\mathrm{SU(2)}$-gauged
  $\mathrm{ST[2,6]}$ model}
\label{sec-gensolST[2,6]}
%%%%%%%%%%%%%%%%%%%%%%%%%%%%%%%%%%%%%%%%%%%%%%%%%%%%%%%%%%%%%%%%%%%%%%
%%%%%%%%%%%%%%%%%%%%%%%%%%%%%%%%%%%%%%%%%%%%%%%%%%%%%%%%%%%%%%%%%%%%%%
%%%%%%%%%%%%%%%%%%%%%%%%%%%%%%%%%%%%%%%%%%%%%%%%%%%%%%%%%%%%%%%%%%%%%%
%%%%%%%%%%%%%%%%%%%%%%%%%%%%%%%%%%%%%%%%%%%%%%%%%%%%%%%%%%%%%%%%%%%%%%

In this section we are going to apply the solution-generating
technique described in the previous section to the particular model we
are interested in: the SU$(2)$-gauged $\rm{ST[2,6]}$ model, which can
be obtained by compactification of Heterotic Supergravity on $T^{4}$
followed by a truncation of all the fields related to the compact
space. The ungauged model has $n_{v}=5$ vector multiplets and is
characterized by a $C_{IJK}$ tensor whose only non-vanishing
components are $C_{0xy}=\tfrac{1}{6}\eta_{xy}$, with $x=1, ...,
5$. The last three vector fields ($x=3,4,5$) will be used as SU$(2)$
gauged fields in the gauged theory.

Then, it is convenient to split the index labelling the vector fields
into a pair of indices $I=\left(i, A+2\right)$, where $i=0,1,2$ labels
the Abelian vector fields and $A=1,2,3$, the SU$(2)$ gauged fields.
Furthermore, we define the following combinations of Abelian vector
fields

\begin{equation}
A^{\pm} \equiv A^{1}\pm A^{2}\, ,
\end{equation}

\noindent
and we are going to use the scalar fields $\phi,k$ and $\ell^{A}$.
The scalar $\phi$ is the string dilaton field; $k$ is the Kaluza-Klein
scalar that measures the size of the circle of the compactification
from 6 to 5 dimensions and the $\ell^{A}$ are just a convenient
SU$(2)$ triplet of scalar fields. The relation between these fields
and those of the standard parametrization in Eq.~(\ref{eq:scalarsp1})
is 

\begin{eqnarray}
  e^{-2\phi}
  & = &
        \tfrac{1}{2}(\phi^{1}-\phi^{2})\, ,
  \\
  \nonumber \\
  k^{4}
  & = &
        {\displaystyle
       2\left[ \frac{\left(\phi^{1}\right)^2-\left(\phi^{2}\right)^2-\phi^{A}\phi^{A}}{\phi^{1}-\phi^{2}}\right]^2\, ,
        }
  \\
  \nonumber \\
  \ell^{A}
  & = &
        \phi^{A}/(\phi^{1}-\phi^{2})\, .
\end{eqnarray}

Then, for this model and using these variables, the generic action
Eq.~(\ref{eq:genericN1d5SEYMaction}) takes the specific form

%\begin{equation}
%\label{eq:gaugedST[2,6]action}
%\begin{array}{rcl}
%S 
%& = &  
%{\displaystyle\int 
%d^{5}x\sqrt{g}\
%\biggl\{
%R
%+\partial_{\mu}\phi\partial^{\mu}\phi
%+\tfrac{4}{3}\partial_{\mu}\log{k}\partial^{\mu}\log{k}
%+2e^{-\phi}k^{-2}\mathfrak{D}_{\mu}\ell^{A}
%\mathfrak{D}^{\mu}\ell^{A}
%}
%\\ \\ & & 
%          -\tfrac{1}{12} e^{2\phi} k^{-4/3}
%          \left(
%          F^{0}\cdot F^{0}
%+\tfrac{1}{2} F^{+}\cdot F^{+}
%+\tfrac{1}{2} F^{-}\cdot F^{-}
%          \right)
%\\ \\ & & 
%          -\tfrac{1}{12}\left(\tfrac{1}{2}e^{-\phi}k^{2/3}
%          -e^{-2\phi}k^{-4/3}\ell^{2}\right) F^{+}\cdot F^{-}
%\\ \\ & & 
%          -\tfrac{1}{12}\left(e^{-\phi}k^{2/3}\delta_{AB}
%+4e^{-2\phi}k^{-4/3}\ell^{A}\ell^{B}\right)
%F^{A}\cdot F^{B}
%\\ \\ & & 
%+\tfrac{1}{24\sqrt{3}}
%{\displaystyle\frac{\varepsilon^{\mu\nu\rho\sigma\alpha}}{\sqrt{g}}}
%          A^{0}{}_{\mu}\left( F^{+}{}_{\nu\rho}F^{-}{}_{\sigma\alpha}
%          -F^{A}{}_{\nu\rho}F^{A}{}_{\sigma\alpha}\right)
%\biggr\}\, .
%\end{array}
%\end{equation}

\begin{equation}
\label{eq:gaugedST[2,6]action}
\begin{aligned}
  S & = \int d^{5}x \, \sqrt{|g|}\, \left\{ R
    +\partial_{\mu}\phi\partial^{\mu}\phi
    +\tfrac{4}{3}\partial_{\mu}\log{k}\partial^{\mu}\log{k}
    +2e^{-\phi}k^{-2}\mathfrak{D}_{\mu}\ell^{A}\mathfrak{D}^{\mu}\ell^{A}
  \right.
  \\
  & \\
  & -\tfrac{1}{12}e^{2\phi}k^{-4/3} F^{0}\cdot F^{0}
  -\tfrac{1}{48}k^{8/3}
  \left(1+2e^{-\phi}k^{-2}\ell^{B}\ell^{B}\right)^{2}F^{+}\cdot F^{+}
  -\tfrac{1}{12}e^{-2\phi}k^{-4/3} F^{-}\cdot F^{-}
  \\
  &\\
  & -\tfrac{1}{6}e^{-2\phi}k^{-4/3}\ell^{B}\ell^{B}F^{+}\cdot F^{-}
  -\tfrac{1}{12}\left(e^{-\phi}k^{2/3}\delta_{AB}
    +4e^{-2\phi}k^{-4/3}\ell^{A} \ell^{B}\right)F^{A}\cdot F^{B}
  \\
  &\\
  &
  -\tfrac{1}{6}e^{-\phi}k^{2/3}\left(1+2e^{-\phi}k^{-2}\ell^B\ell^B\right)\ell^{A}
  F^{+}\cdot F^{A} -\tfrac{1}{3}
  e^{-2\phi}k^{-4/3}\ell^{A} F^{-}\cdot  F^{A}
  \\
  &\\
  & \left.+\tfrac{1}{24\sqrt{3}}
    {\displaystyle\frac{\varepsilon^{\mu\nu\rho\sigma\alpha}}{\sqrt{|g|}}}
    A^{0}{}_{\mu}\left( F^{+}{}_{\nu\rho}F^{-}{}_{\sigma\alpha}
      -F^{A}{}_{\nu\rho}F^{A}{}_{\sigma\alpha}\right)\right\}\, .
\end{aligned}
\end{equation}

Following the redefinition of the vector fields, in order to describe
the construction of the solutions, we will use the functions

\begin{equation}
Z_{\pm}\equiv Z_{1}\pm Z_{2}\, ,
  \,\,\,\,\, \text{and}\,\,\,\,\,
\tilde{Z}_{+}=Z_{+}-Z_{A} Z_{A}/Z_{-}\, .
\end{equation}

%%%%%%%%%%%%%%%%%%%%%%%%%%%%%%%%%%%%%%%%%%%%%%%%%%%%%%%%%%%%%%%%%%%%%%
%%%%%%%%%%%%%%%%%%%%%%%%%%%%%%%%%%%%%%%%%%%%%%%%%%%%%%%%%%%%%%%%%%%%%%
\subsection{The metric}
%%%%%%%%%%%%%%%%%%%%%%%%%%%%%%%%%%%%%%%%%%%%%%%%%%%%%%%%%%%%%%%%%%%%%%
%%%%%%%%%%%%%%%%%%%%%%%%%%%%%%%%%%%%%%%%%%%%%%%%%%%%%%%%%%%%%%%%%%%%%%

According to the prescription given in the previous section, the
metric of the solutions of this model that we can construct with it
will have the general form Eq.~(\ref{eq:gen-metric}), but now with the
metric function $f$ taking the form

\begin{equation}
  \label{eq:fmetricfunction}
f^{-3}=\tfrac{27}{2}Z_{0} \tilde{Z}_{+}Z_{-}\, .
\end{equation}

%%%%%%%%%%%%%%%%%%%%%%%%%%%%%%%%%%%%%%%%%%%%%%%%%%%%%%%%%%%%%%%%%%%%%%
%%%%%%%%%%%%%%%%%%%%%%%%%%%%%%%%%%%%%%%%%%%%%%%%%%%%%%%%%%%%%%%%%%%%%%
\subsection{The scalar fields}
%%%%%%%%%%%%%%%%%%%%%%%%%%%%%%%%%%%%%%%%%%%%%%%%%%%%%%%%%%%%%%%%%%%%%%
%%%%%%%%%%%%%%%%%%%%%%%%%%%%%%%%%%%%%%%%%%%%%%%%%%%%%%%%%%%%%%%%%%%%%%

In terms of the functions that we have defined, the scalars are given
by

\begin{equation}
  \label{eq:physcalarsp2}
  e^{2\phi}=2\frac{Z_{0}}{Z_{-}} \, ,
  \hspace{1.5cm}
  k=\left(\frac{2\tilde{Z}_{+}^{2}} {Z_{0}Z_{-}} \right)^{1/4}\, ,
  \hspace{1.5cm}
  \ell^{A}=\frac{Z_{A}}{Z_{-}}\, .
\end{equation}

%%%%%%%%%%%%%%%%%%%%%%%%%%%%%%%%%%%%%%%%%%%%%%%%%%%%%%%%%%%%%%%%%%%%%%
%%%%%%%%%%%%%%%%%%%%%%%%%%%%%%%%%%%%%%%%%%%%%%%%%%%%%%%%%%%%%%%%%%%%%%
\subsection{The vector fields}
%%%%%%%%%%%%%%%%%%%%%%%%%%%%%%%%%%%%%%%%%%%%%%%%%%%%%%%%%%%%%%%%%%%%%%
%%%%%%%%%%%%%%%%%%%%%%%%%%%%%%%%%%%%%%%%%%%%%%%%%%%%%%%%%%%%%%%%%%%%%%

The vector fields are generically given by Eq.~(\ref{eq:gen-vectors}),
but here we will restrict ourselves to solutions with
$\hat{A}^{0}=\hat{A}^{\pm}=0$ for simplicity. We will keep the $\hat{A}^{A}\neq 0$, though. Then,
the vector fields of our solutions will have the form

\begin{eqnarray}
  A^{0}
  & = &
        -\frac{1}{\sqrt{3}}\frac{1}{Z_{0}}\left(dt+ \omega\right)\, ,
  \\
  & & \nonumber \\
  A^{\pm}
  & = &
        -\frac{2}{\sqrt{3}}\frac{Z_{+}}{Z_{\pm} \tilde{Z}_{+}}
        \left(dt+ \omega\right)\, ,
  \\
    & & \nonumber \\
  A^{A}
  & = &
        \frac{2}{\sqrt{3}}\frac{Z_{A}}{\tilde{Z}_{+}Z_{-}}
        \left(dt+ \omega\right)+\hat{A}^{A}\, ,
\end{eqnarray}

\noindent
and the building blocks for which we will have to solve
Eqs.~(\ref{eq:selfduality}), (\ref{eq:Zs}) and (\ref{eq:omega}) are
the functions $Z_{0},Z_{\pm},Z_{A}$ and the 1-forms
$\omega, \hat{A}^{A}$.

Let us first consider Eq.~(\ref{eq:selfduality}). This equation can be
solved in an arbitrary HK metric by SU$(2)$ gauge fields
$\hat{A}^{A}$ defined on it via a \textit{generalized 't Hooft ansatz}
\cite{Chimento:2018kop}:

\begin{equation}
\label{eq:thooft}
g\hat{A}^{A}=\bar{\eta}^{A}{}_{mn} \partial_{n}\log{P} \, v^{m}\, ,
\end{equation}

\noindent
where $P$ is a harmonic function in the HK space, the
$\bar{\eta}^{A}{}_{mn}$ are the three antiselfdual complex structures
that are covariantly conserved in the HK space and which
can be taken to be the constant 't~Hooft symbols\footnote{The 't~Hooft
  symbols satisfy the following identities
  \begin{eqnarray}
  \epsilon^{ABC}\bar{\eta}^{B}{}_{mp}\bar{\eta}^{C}{}_{nq}
  & = &
  -\delta_{mn}\bar{\eta}^{A}{}_{pq}
    -\delta_{pq}\bar{\eta}^{A}{}_{mn}
    +\delta_{mq}\bar{\eta}^{A}{}_{pn}+\delta_{pn}\bar{\eta}^{A}{}_{mq}\, ,
  \\
  \bar{\eta}^{A}{}_{mn}\bar{\eta}^{A}{}_{pq}
  & = &
  2\delta_{m[p}\delta_{q]n}-\epsilon_{mnpq}\, ,
  \\
  \bar{\eta}^{A}{}_{mp}\bar{\eta}^{B}{}_{pn}
  & = &
  -\delta^{AB}\delta_{mn}+\epsilon^{ABC}\bar{\eta}^{C}{}_{mn}\, .
\end{eqnarray}
}. Finally, the $v^{m}$ are the vierbein of the HK space:
$d\sigma^{2}=v^{m} v^{m}$.

The selfdual gauge field strength is given by\footnote{The 
  gauge field strength for the SU$(2)$ group is given by
  \begin{equation}
  F^{A}=dA^{A}+ \tfrac{1}{2}\epsilon^{ABC}\, A^{B} \wedge A^{C}\, ,  
\end{equation}
and the covariant derivatives of the scalar functions are given by
\begin{eqnarray}
  \mathfrak{D}Z_{A}
  & = &
        dZ_{A} -\epsilon^{ABC}A^{B}Z_{C}\, ,
  \\
  & & \nonumber \\
  \mathfrak{D}\varphi^{A}
  & = &
        d\varphi^{A} +\epsilon^{ABC}A^{B}\varphi^{C}\, .
\end{eqnarray}
}

\begin{equation}
  \label{eq:FAsthooft}
  g\hat{F}^{A}
  =
  \left[\bar{\eta}^{A}{}_{np}\nabla_{m}\partial_{p} \log{P}
    +\bar{\eta}^{A}{}_{mp}\,\partial_{p}\log{P}\,\partial_{n}\log{P}
    -\tfrac{1}{2}\bar{\eta}^{A}{}_{mn}\left(\partial \log{P}\right)^{2}\right]
  v^{m}\wedge v^{n}\, .
\end{equation}

Next, let us focus on Eqs.~(\ref{eq:Zs}), which in this case take the
form

\begin{eqnarray}
\label{eq:Z0}
d\star_{\sigma} d Z_{0} & = &  \frac{1}{18}\hat{F}^{A} \wedge \hat{F}^{A}\, ,\\
\nonumber\\
\label{eq:Z12}
d\star_{\sigma} d Z_{1,2} & = & 0\, ,\\
\nonumber\\
\label{eq:ZA}
\hat{\mathfrak{D}} \star_{\sigma} \hat{\mathfrak{D}} Z_{A} & = & 0\, .
\end{eqnarray}

For the configurations considered, it was shown in
Ref.~\cite{Chimento:2018kop} that the instanton number density
$\hat{F}^{A} \wedge \hat{F}^{A}$ enjoys the so-called ``Laplacian
property'', \textit{i.e.}

\begin{equation}
  \hat{F}^{A}\wedge \hat{F}^{A}
  =
  -d\star_{\sigma} d\left[\frac{\left(\partial \log{P}\right)^{2}}{g^{2}}\right]
  \sqrt{h}\: d^{4}x\,,
\end{equation}

\noindent
where $h$ is the determinant of the HK metric.
Hence, we find that Eqs.~(\ref{eq:Z0}) and (\ref{eq:Z12}) are solved by 

\begin{eqnarray}
Z_{0} & = & Z^{(0)}_{0} -\frac{\left(\partial \log{P}\right)^{2}}{18g^{2}} \, ,\\
\nonumber\\
Z_{1,2} & = & Z^{(0)}_{1,2}\, .
\end{eqnarray}

\noindent
where $Z^{(0)}_{0,1,2}$ are  harmonic functions on the HK space.

As for Eq.~(\ref{eq:ZA}), two solutions of it are known to us:

%%%%%%%%%%%%%%%%%%%%%%%%%%%%%%%%%%%%%%%%%%%%%%%%%%%%%%%%%%%%%%%%%%%%%%
%%%%%%%%%%%%%%%%%%%%%%%%%%%%%%%%%%%%%%%%%%%%%%%%%%%%%%%%%%%%%%%%%%%%%%
\subsubsection{Solution D1}
%%%%%%%%%%%%%%%%%%%%%%%%%%%%%%%%%%%%%%%%%%%%%%%%%%%%%%%%%%%%%%%%%%%%%% 
%%%%%%%%%%%%%%%%%%%%%%%%%%%%%%%%%%%%%%%%%%%%%%%%%%%%%%%%%%%%%%%%%%%%%%

This solution was found in Ref.~\cite{Ramirez:2016tqc} for HK metrics
admitting a triholomorphic isometry. These metrics are known as
Gibbons-Hawking (GH) metrics \cite{Gibbons:1979zt,Gibbons:1987sp} and
can be put in the form

  \begin{equation}
    \label{eq:GHmetric}
    d\sigma^{2}= H^{-1}(d\eta +\chi)^{2}+H\,d x^{i} d x^{i} \, ,
    \qquad
    d\chi=\star_{(3)} \, dH\, ,
\end{equation}

\noindent
where $\star_{(3)}$ is the Hodge star on $\mathbb{E}^{3}$ and $H$ is a
($\eta$-independent) harmonic function on
$\mathbb{E}^{3}$.\footnote{This is the integrability condition of the
  equation $d\chi=\star_{(3)} \, dH$.}

This solution also makes use of SU$(2)$ instantons obtained through
the 't~Hooft anstaz Eq.~(\ref{eq:thooft}) with a function $P$ which is
also independent of $\eta$ and, therefore, harmonic on
$\mathbb{E}^{3}$ as well. The consistency of the solution demands that
the functions $Z_{A}$ are also independent of $\eta$.

Let us see in detail how this solution is obtained.

If $P$ is independent of $\eta$ and the HK metric is the
above GH metric, the vector fields defined by `t~Hooft
ansatz Eq.~(\ref{eq:thooft}) can be written in the simple form

\begin{equation}
  \label{eq:thooftiso}
  g \hat{A}^{A}
  =
  H^{-1}\varphi^{A} \left(d\eta+\chi\right)+\breve{A}^{A}\, ,
\end{equation}

\noindent
where $\varphi^{A}$ and $\breve{A}^{A}$ are fields (scalar and vector,
resp.) defined on $\mathbb{E}^{3}$ and determined by the choice of $P$
by

\begin{eqnarray}
\label{eq:varphiA}  
\varphi^{A} & = & \delta^{A i} \, \partial_{\underline{i}}\log{P}\, , \\
\nonumber\\
\breve{A}^{A} & = & \epsilon^{Aij}\,\partial_{\underline{i}}\log{P}dx^{j}\, .
\end{eqnarray}

The selfduality of the field strength of $\hat{A}^{A}$ reduces in this
case to the Bogmol'nyi equation relating the field strength of
$\breve{A}^{A}$ and the covariant derivative of the $\varphi^{A}$ with
respect to that connection on $\mathbb{E}^{3}$:

\begin{equation}
  \label{eq:Bogomolnyi}
\star_{(3)} \breve{F}^{A}=-\breve{\mathfrak{D}}\varphi^{A}\, .
\end{equation}

Substituting Eq.~(\ref{eq:thooftiso}) into Eq.~(\ref{eq:ZA}), we find
the following equation for $Z_{A}$

\begin{equation}
  \label{eq:ZAiso}
  \partial_{\underline{i}}\partial_{\underline{i}} Z_{A}
  +2\delta_{A}^{i}\,\partial_{\underline{i}} Z_{B} \,\varphi^{B}
  -2\varphi^{A}\delta^{Bj} \partial_{\underline{j}} Z_{B}
  -2 Z_{A} \,\varphi^{B}\varphi^{B}=0\, .
\end{equation}

Following Ref.~\cite{Ramirez:2016tqc}, we make the following ansatz for $Z_{A}$:

\begin{equation}\label{eq:sol1}
 Z_{A}=\delta_{A}^{i}\frac{\partial_{\underline{i}} Q}{g P}\, ,
\end{equation}

\noindent
where $Q$ is some function on $\mathbb{E}^{3}$.  Plugging this ansatz
into Eq.~(\ref{eq:ZAiso}), we arrive at the following equation for $Q$:

\begin{equation}
  P \partial_{\underline{i}}
  \left(\frac{\partial_{\underline{j}}\partial_{\underline{j}}Q}{gP^{2}}\right)
  =
  0\, ,
  \qquad
  \Rightarrow
  \qquad
  \partial_{\underline{j}}\partial_{\underline{j}}Q=k P^{2}\, ,
\end{equation}

\noindent
for some constant $k$. If, as we will assume later,
$P=1+\lambda^{-2}/r$ (a spherically-symmetric harmonic function in
$\mathbb{E}^{3}$), then

\begin{equation}
Q = Q^{(0)}+ k\left[ \tfrac{1}{6}r^{2}+\lambda^{-2}r +\lambda^{-4}\log{r} \right]\, ,   
\end{equation}

\noindent
where $Q^{(0)}$ is another harmonic function in
$\mathbb{E}^{3}$. $k\neq 0$ will, in general, give rise to
non-asymptotically flat metrics and, therefore, we will set it to
zero.

%%%%%%%%%%%%%%%%%%%%%%%%%%%%%%%%%%%%%%%%%%%%%%%%%%%%%%%%%%%%%%%%%%%%%%
%%%%%%%%%%%%%%%%%%%%%%%%%%%%%%%%%%%%%%%%%%%%%%%%%%%%%%%%%%%%%%%%%%%%%%
\subsubsection{Solution D2}
%%%%%%%%%%%%%%%%%%%%%%%%%%%%%%%%%%%%%%%%%%%%%%%%%%%%%%%%%%%%%%%%%%%%%% 
%%%%%%%%%%%%%%%%%%%%%%%%%%%%%%%%%%%%%%%%%%%%%%%%%%%%%%%%%%%%%%%%%%%%%%
  
The second solution available in the literature was found in
Ref.~\cite{Lambert:1999ua} in $\mathbb{E}^{4}$ and it was used to
construct a dyonic instanton solution of Heterotic supergravity in
\cite{Eyras:2000dg}. The generalization of this solution to the case
of arbitrary HK metrics is straightforward. In order to show this, let
us first rewrite Eq.~(\ref{eq:ZA}) as

\begin{equation}
  \label{eq:ZAexpanded}
\begin{aligned}
  d\star_{\sigma} d Z_{A}- g\,\epsilon^{ABC} \,Z_{B}\, d\star_{\sigma}
  \hat{A}^{C}
  -2 g\,\epsilon^{ABC}(\star_{\sigma} \hat{A}^{B}) \wedge dZ_{C}
  &
\\
 +g^{2}\left(Z_{B}\hat{A}^{B}\wedge(\star_{\sigma} \hat{A}^{A})
  -Z_{A}\,\hat{A}^{B}\wedge (\star_{\sigma} \hat{A}^{B})\right)
 = 0\, .
\end{aligned}
\end{equation}

Let us keep just one of the $Z_{A}$ functions active, say
$Z_{3}$. Substituting the 't~Hooft ansatz Eq.~(\ref{eq:thooft}) into
Eq.~(\ref{eq:ZAexpanded}), we get the following conditions

\begin{eqnarray}
  \bar{\eta}^{1}{}_{mn}\partial_{n} P \,\partial_{m} Z_{3}
  =
  \bar{\eta}^{2}{}_{mn}\partial_{n} P \,\partial_{m} Z_{3}
  & = &
        0\, , \\
& &   \nonumber\\
  \nabla^{2} Z_{3}-2\, Z_{3}\, \left(\partial\log{P}\right)^{2}
  & = &
        0\, ,
\end{eqnarray}

\noindent
which are solved by

\begin{equation}
  \label{eq:sol2}
Z_{3}=\frac{\xi_{2}}{gP}\, ,
\end{equation}

\noindent
where $\xi_{2}$ is an arbitrary constant. 

Notice that when the HK metric is a GH metric and the vector fields do
not depend on the isometric coordinate $\eta$, Eq.~(\ref{eq:sol2}) is
a particular case of Eq.~(\ref{eq:sol1}) given by the choice of
harmonic function $Q(\vec x)=\xi_{2} \,x^{3}$.

%%%%%%%%%%%%%%%%%%%%%%%%%%%%%%%%%%%%%%%%%%%%%%%%%%%%%%%%%%%%%%%%%%%%%%
%%%%%%%%%%%%%%%%%%%%%%%%%%%%%%%%%%%%%%%%%%%%%%%%%%%%%%%%%%%%%%%%%%%%%%
%%%%%%%%%%%%%%%%%%%%%%%%%%%%%%%%%%%%%%%%%%%%%%%%%%%%%%%%%%%%%%%%%%%%%%
%%%%%%%%%%%%%%%%%%%%%%%%%%%%%%%%%%%%%%%%%%%%%%%%%%%%%%%%%%%%%%%%%%%%%%
\subsection{The 1-form $\omega$}
%%%%%%%%%%%%%%%%%%%%%%%%%%%%%%%%%%%%%%%%%%%%%%%%%%%%%%%%%%%%%%%%%%%%%%
%%%%%%%%%%%%%%%%%%%%%%%%%%%%%%%%%%%%%%%%%%%%%%%%%%%%%%%%%%%%%%%%%%%%%%
%%%%%%%%%%%%%%%%%%%%%%%%%%%%%%%%%%%%%%%%%%%%%%%%%%%%%%%%%%%%%%%%%%%%%%
%%%%%%%%%%%%%%%%%%%%%%%%%%%%%%%%%%%%%%%%%%%%%%%%%%%%%%%%%%%%%%%%%%%%%%

Finally, let us consider Eq.~(\ref{eq:omega}). When the HK
metric is a GH metric of the form
Eq.~(\ref{eq:GHmetric}), we can always write the 1-form $\omega$ as

\begin{equation}
\label{eq:sol2omega}
\omega=\omega_{5} (d\eta+\chi)+\breve \omega\, ,
\end{equation}

\noindent
and then Eq.~(\ref{eq:omega}) takes the following form:

\begin{equation}
  \label{eq:omegaiso}
\begin{aligned}
  d\breve \omega+\omega_{5} d\chi-H\star_{(3)} d\omega_{5}
  & =
  \frac{\sqrt{3}}{g}
  \left(\frac{Z_{B}\varphi^{B}}{H}d\chi+Z_{B} \breve{F}^{B}\right)
  \\
  \\
  & =
  \frac{\sqrt{3}}{2g}
  \left[
    \frac{Z_{B}\varphi^{B}}{H}d\chi
    -H \star_{(3)} d\left(\frac{Z_{B}\varphi^{B}}{H}\right)
  \right.
  \\
  \\
  &
  \,\,\,\,\,\,
  \left.
    +\star_{(3)}\left(\varphi^{B} \:\mathfrak {\breve D} Z_{B}
      -Z_{B} \mathfrak {\breve D}\varphi^{B}\right)\right]\, ,
\end{aligned}
\end{equation}

\noindent
where we have made use of the Bogomol'nyi equation
(\ref{eq:Bogomolnyi}) in order to rewrite the r.h.s.~of the
equation. The integrability condition of this equation is\footnote{To
  derive this equation, we use that
\begin{equation}
  d\star_{(3)} \left(\varphi^{B} \:\mathfrak {\breve D} Z_{B}
    -Z_{B} \mathfrak {\breve D}\varphi^{B}\right)
  =
  \varphi^{B} \, \breve {\mathfrak{D}}\star_{(3)} \breve {\mathfrak{D}} Z_{B}
  =
  0\, ,
\end{equation}
as a consequence of the Bogomol'nyi equation (\ref{eq:Bogomolnyi}).
} 

\begin{equation}
  H\, d\star_{(3)} d
  \left(\omega_{5}-\frac{\sqrt{3}}{2g}\frac{Z_{B}\varphi^{B}}{H}\right)
  =
  0\, ,
\end{equation}

\noindent
so that

\begin{equation}
  \label{eq:omega5}
  \omega_{5}
  =
  M+\frac{\sqrt{3}}{2g}\frac{ Z_{B}\varphi^{B}}{H }\, ,
\end{equation}

\noindent
where $M$ is another harmonic function in $\mathbb{E}^{3}$. Finally,
substituting Eq.~(\ref{eq:omega5}) back in Eq.~(\ref{eq:omegaiso}), we
arrive at the following equation for $\omega$:

\begin{equation}
  \label{eq:breveomega}
  \star_{(3)} d\breve\omega
  =
  HdM-MdH
  +\frac{\sqrt{3}}{2g}\left(\varphi^{B} \:\mathfrak {\breve D} Z_{B}
    -Z_{B} \mathfrak {\breve D}\varphi^{B}\right)\, .
\end{equation}

This is the equation that will have to be solved if we use the D1
solution for the gauge fields or if we use the D2 solution over a
GH space.

For the D2 solution Eq.~(\ref{eq:sol2}) over a generic HK
space, it is natural to try an ansatz of the form

\begin{equation}\label{eq:ansatzomega}
\omega = \bar{\eta}^{3}{}_{mn}\partial_{n} \Omega \, v^{m}\, .
\end{equation}

Then, using Eqs.~(\ref{eq:FAsthooft}) and (\ref{eq:ansatzomega}), we
find  that Eq.~(\ref{eq:omega}) reduces to

\begin{equation}
\begin{aligned}
  2\,\bar\eta^{3}{}_{[n|p}\nabla_{m]} \partial_{p}\Omega
  +\tfrac{1}{2}\bar\eta^{3}{}_{mn}\nabla^{2}\Omega
  &
  =
  \frac{\sqrt{3}\xi}{g^{2} P}
  \left\{\bar{\eta}^{3}{}_{[n|p}\nabla_{m]}\partial_{p} \log{P}\right.
  \\
  \\
  &
  \,\,\,\,\,\,
  \left.
    +\bar{\eta}^{3}{}_{[m|p}\,\partial_{p}\log{P}\,\partial_{n]}\log{P}
    -\tfrac{1}{2}\bar{\eta}^{3}{}_{mn}\left(\partial \log{P}\right)^{2}\right\}\, ,
\end{aligned}
\end{equation}

\noindent
which is solved by 

\begin{equation}
\Omega=-\frac{\sqrt{3}\xi_{2}}{2g^{2}} P^{-1}\, .
\end{equation}

Summarizing, for the solutions D1 and D2 of the gauge fields, the
1-form $\omega$ is given by

\begin{equation}
  \label{eq:generalsolomega}
\begin{aligned}
  \text{D1 \& D2 on GH space:}\qquad \omega
  & =
  \left(M+\frac{\sqrt{3}}{2g}\frac{ Z_{B}\, \varphi^{B}}{H}\right)
  \left(d\eta+\chi\right)+\breve{\omega}\, ,
  \\
\\
\text{D2 on general HK space:}\qquad\omega
& =
\frac{\sqrt{3}\, \xi_{2}}{2 g^{2} P^{2}}\,\bar{\eta}^{3}{}_{mn}\,
\partial_{n} P\, v^{m}\, .
\end{aligned}
\end{equation}

\noindent
with $\breve{\omega}$ satisfying Eq.~(\ref{eq:breveomega}). 

%%%%%%%%%%%%%%%%%%%%%%%%%%%%%%%%%%%%%%%%%%%%%%%%%%%%%%%%%%%%%%%%%%%%%%
%%%%%%%%%%%%%%%%%%%%%%%%%%%%%%%%%%%%%%%%%%%%%%%%%%%%%%%%%%%%%%%%%%%%%%
\subsection{Dimensional reduction to $d=4$}
\label{sec:dimensionalreduction}
%%%%%%%%%%%%%%%%%%%%%%%%%%%%%%%%%%%%%%%%%%%%%%%%%%%%%%%%%%%%%%%%%%%%%%
%%%%%%%%%%%%%%%%%%%%%%%%%%%%%%%%%%%%%%%%%%%%%%%%%%%%%%%%%%%%%%%%%%%%%%

When the HK metric is a GH metric taking the form
Eq.~(\ref{eq:GHmetric}) in the coordinate system adapted to the
triholomorphic isometry and (quite naturally) none of the physical
fields depends on the isometric coordinate $\eta$, it is possible to
perform a standard Kaluza-Klein reduction of the solution to $d=4$
along that direction to obtain a solution of the SU$(2)$-gauged
$\rm{ST[2,6]}$ model of $\mathcal{N}=2,d=4$ SEYM. The matter fields of
this theory are vector fields $A^{\Lambda}{}_{\mu}$,
$\Lambda=0,1,\cdots,6$ and complex scalars $Z^{i}$, $i=1,\ldots,6$
parametrizing the coset space

\begin{equation}
  \frac{\mathrm{SL}(2,\mathbb{R})}{\mathrm{SO}(2)}\times
  \frac{\mathrm{SO}(2,5)}{\mathrm{SO}(2)\times \mathrm{SO}(5)}\, .
\end{equation}

The interactions are determined by the cubic prepotential

\begin{equation}
  \mathcal{F}
  =
  -\tfrac{1}{3!}
  \frac{d_{ijk}\mathcal{X}^{i}\mathcal{X}^{j}\mathcal{X}^{k}}{\mathcal{X}^{0}}\, ,
\end{equation}

\noindent
where the constant, fully symmetric, tensor $d_{ijk}$ is related to
the tensor $C_{IJK}$ of the 5-dimensional theory by

\begin{equation}
  d_{ijk}=6C_{i-1\, j-1\, k-1}\, ,
  \hspace{1cm}
  i,j,k=1,\ldots,6\, .
\end{equation}
The index $1$ corresponds to the 5-dimensional $0$ and the
4-dimensional $0$ is associated to the Kaluza-Klein vector of the 
dimensional reduction fomr 5 to 4 dimensions.

In this model, the SU$(2)$ gauge group acts on the complex scalars and
vector fields with indices $4,5,6$. Furthermore, the $+$ and $-$
combinations defined in the 5-dimensional case now correspond to

\begin{equation}
A^{\pm}{}_{\mu}\equiv A^{2}{}_{\mu}\pm A^{3}{}_{\mu}\, .  
\end{equation}

A solution-generating technique to construct directly the timelike
supersymmetric solutions of $\mathcal{N}=2,d=4$ SEYM theories was
found in Refs.~\cite{Huebscher:2007hj,Hubscher:2008yz,Meessen:2012sr},
but the procedure turns out to be completely equivalent to the
construction of timelike supersymmetric solutions with an additional
triholomorphic isometry in the auxiliary HK space (sometimes called
\textit{base space}) in $d=5$.

The formulae relating the 4- and 5-dimensional fields were given in
full generality in Ref.~\cite{Meessen:2015enl}. Here, we particularize
those formulae for the dyonic configurations considered in this
paper.\footnote{It is worth mentioning that there are many purely
  magnetic solutions of the Bogomol'nyi equations (and, hence, of the
  selfduality equations) which are, by definition, non-Abelian BPS
  magnetic monopoles. They were found by Protogenov in
  Ref.~\cite{Protogenov:1977tq} and all of them can and have been used
  to construct regular 4-dimensional black-hole solutions with
  non-Abelian fields in these theories
  \cite{Huebscher:2007hj,Meessen:2008kb,Hubscher:2008yz,Meessen:2012sr,Meessen:2015enl}. However,
  not all these magnetic monopoles correspond to regular instantons in
  5 dimensions and, therefore, they cannot be used to construct
  regular 5-dimensional black holes. Here we are solving directly the
  selfduality equation (\ref{eq:selfduality}) and it is guaranteed (it is
  tautological) that, if they are defined on a GH space
  and are independent of $\eta$, they give solutions of the
  Bogomol'nyi equations which correspond to regular instantons.}

We find that the 4-dimensional Einstein-frame metric takes the
standard form of the timelike supersymmetric solutions of
$\mathcal{N}=2,d=4$ SEYM theories

\begin{equation}
  ds_{(4)}^{2}
  =
  e^{2U}\left(dt+\breve{\omega}\right)^{2}-e^{-2U}\, dx^{i} dx^{i}\, ,
\end{equation}

\noindent
where the metric function $e^{-2U}$ is given by

\begin{equation}
  e^{-2U}
  =
  \sqrt{f^{-3} H-\left(\omega_{5} H\right)^{2}}
  =
  \sqrt{\frac{27}{2}Z_{0}\tilde{Z}_{+}Z_{-} H-\left(\omega_{5} H\right)^{2}}\, ,
\end{equation}

\noindent
and the 1-form $\breve{\omega}$ is the solution to
Eq.~(\ref{eq:breveomega}).

The 4-dimensional vector fields are given by\footnote{We have added a
  subindex $_{(4)}$ to distinguish them from the 5-dimensional ones.}

\begin{eqnarray}
  \label{eq:4d-AKK}
  A^{0}_{(4)}
  & = &
        \frac{1}{2\sqrt{2}}\,
        \left[
        -e^{4U}H^{2} \omega_{5} \left(dt+\breve{\omega}\right)+ \chi
        \right]\, ,
  \\
\nonumber\\
\label{eq:4d-A0}
  A^{1}_{(4)}
  & = &
        \frac{1}{6\sqrt{2}}
        \frac{e^{4U} H f^{-3}}{Z_{0}} \, \left(dt+\breve{\omega}\right)\, ,
  \\
\nonumber\\
\label{eq:4d-Apm}
  A^{\pm}_{(4)}
  & = &
        \frac{1}{3\sqrt{2}}\frac{e^{4U} H f^{-3}Z_{+}}{ Z_{\pm} \tilde{Z}_{+}}\,
        \left(dt+\breve{\omega}\right)\, ,
  \\
\nonumber\\
  A^{A}_{(4)}
  & = &
        \frac{-e^{4U}H}{3\sqrt{2}}
        \left(\frac{ f^{-3}Z_{A}}{\tilde{Z}_{+} Z_{-}}
        +\frac{\sqrt{3}\omega_{5} \varphi^{A}}{2g}\right)
        \left(dt+\breve\omega\right)+\frac{1}{g}\,\breve{A}^{A}\, ,
\end{eqnarray}

\noindent
where the $\varphi^{A}$ are defined in Eq.~(\ref{eq:varphiA}).

Finally, the six  4-dimensional scalars are given by\footnote{Note that we
use superscripts to label the 4-dimensional scalars to distinguish them
from the $Z$-functions  which instead have subindices.}

\begin{eqnarray}
  \label{eq:4dscalar0}
  Z^{1}
  & = &
        -\frac{1}{3\, Z_{0}\, H}
        \left(\omega_{5} H-i e^{-2U}\right)\, ,
  \\
  \nonumber\\
   \label{eq:4dscalar+}
  Z^{+}
  & = &
        -\frac{2}{3\,\tilde{Z}_{+}\, H}
        \left(\omega_{5} H-i e^{-2U}\right)\, ,
  \\
  \nonumber\\
  \label{eq:4dscalar-}
  Z^{-}
  & = &
        -\frac{2\, Z_{+}}{3\,\tilde{Z}_{+}\,Z_{-}\, H}
        \left(\omega_{5} H-i e^{-2U}\right)\, ,
  \\
  \nonumber\\
  \label{eq:4dscalarA}
  Z^{A}
  & = &
        \frac{2\, Z_{A}}{3\,\tilde{Z}_{+}\,Z_{-}\, H}
        \left(\omega_{5} H-ie^{-2U}\right)+\frac{\varphi^{A}}{g \,H}\, .
\end{eqnarray}

The Kaluza-Klein scalar of the $5\rightarrow 4$ compactification is a
particular combination of these 6 complex scalars and it is given by 

\begin{equation}
\label{eq:kkscalar5-4}
\ell^2  =  e^{-4U}H^{-2}f^{2}\, .
\end{equation}

%%%%%%%%%%%%%%%%%%%%%%%%%%%%%%%%%%%%%%%%%%%%%%%%%%%%%%%%%%%%%%%%%%%%%% 
%%%%%%%%%%%%%%%%%%%%%%%%%%%%%%%%%%%%%%%%%%%%%%%%%%%%%%%%%%%%%%%%%%%%%%
%%%%%%%%%%%%%%%%%%%%%%%%%%%%%%%%%%%%%%%%%%%%%%%%%%%%%%%%%%%%%%%%%%%%%%
%%%%%%%%%%%%%%%%%%%%%%%%%%%%%%%%%%%%%%%%%%%%%%%%%%%%%%%%%%%%%%%%%%%%%%
\section{Rotating black holes}
\label{explicit-solutions}
%%%%%%%%%%%%%%%%%%%%%%%%%%%%%%%%%%%%%%%%%%%%%%%%%%%%%%%%%%%%%%%%%%%%%%
%%%%%%%%%%%%%%%%%%%%%%%%%%%%%%%%%%%%%%%%%%%%%%%%%%%%%%%%%%%%%%%%%%%%%%
%%%%%%%%%%%%%%%%%%%%%%%%%%%%%%%%%%%%%%%%%%%%%%%%%%%%%%%%%%%%%%%%%%%%%%
%%%%%%%%%%%%%%%%%%%%%%%%%%%%%%%%%%%%%%%%%%%%%%%%%%%%%%%%%%%%%%%%%%%%%%

In the previous section, out of the many solutions that can be
obtained by using the techniques explained in Section~\ref{sec-setup},
we have selected two more restricted classes characterized by dyonic
Yang-Mills fields of two different kinds that we have labeled D1 and
D2. These two classes of solutions still depend on functions and
building blocks which must satisfy certain differential equations and
conditions which do not determine them completely. In this section we
are going to make some particular choices of these building blocks
adequate to find single-center black-hole solutions.\footnote{Multicenter solutions have been considered in Ref.~\cite{Meessen:2017rwm}.}
First of all, although we can use any HK metric for the solutions of
Sec.~\ref{sec-gensolST[2,6]} based on the dyonic instanton D2, we are
going to restrict ourselves to GH metrics, Eq.~(\ref{eq:GHmetric}),
and, in particular, to spherically-symmetric GH metrics of the form

\begin{equation}
  \label{eq:sphericallysymmetricGHmetric}
  d\sigma^{2}
  =
  H^{-1}\left(d\eta+\chi\right)^{2}+H\, \left(dr^{2}+r^{2} d\Omega^{2}_{(2)}\right)\, ,
  \,\,\,\,
  r^{2}=x^{i} x^{i}\, ,
  \,\,\,\,
  d\Omega^{2}_{(2)}=d\theta^{2}+\sin{\theta}^{2}d\phi^{2}\, ,
\end{equation}

\noindent
where $H$ only depends on the radial coordinate $r$ and where the
coordinate $\eta$ is compact and has period
$\eta\sim \eta+2\pi \ell_{s}$, $\ell_{s}$ being a length scale that we
take to be the string length.

Since $H$ is a function of $r$ harmonic in $\mathbb{E}^{3}$, the most
general choice of $H$ and the corresponding $\chi$ are locally given
by

\begin{equation}
H=a_{H}+\frac{b_{H}}{r}\, , \qquad  \chi=b_{H} \cos{\theta} \, d\phi\, ,
\end{equation}

\noindent
where $a_{H},b_{H}$ are two integration constants to be determined.

As it stands, for generic values of $b_{H}$, this metric has an
undesirable feature: it has a Dirac-Misner string.  Fortunately, it
can be eliminated from the metric
(\ref{eq:sphericallysymmetricGHmetric}) by taking

\begin{equation}
b_{H}=n\ell_{s}/2\, , \qquad n\in \mathbb Z\, ,
\end{equation}

\noindent
and covering the HK manifold with two patches. For the time
being, though, we will just study the metric locally in one of those
patches.

Furthermore, when $a_{H}=0$, the compactification to 4 dimensions is 
singular (observe that the KK scalar in (\ref{eq:kkscalar5-4})  blows up), which means 
that the solutions with $a_{H}=0$ only makes sense in 5 dimensions. When $a_{H}\neq0$, 
however, the asymptotic radius of the internal direction is finite and therefore 
 the solution is effectively 4-dimensional. We will deal with these two possibilities separately.

Once the HK metric has been chosen, we have to specify the
magnetic part of the non-Abelian vector fields, $\hat{A}^{A}$, which
is given in terms of the harmonic function $P$ of the `t~Hooft ansatz
by Eq.~(\ref{eq:thooft}). Again, if $P$ depends only on the radial
coordinate, it must be given by

\begin{equation}
P=1+\frac{\lambda^{-2}}{r}\, ,
\end{equation}

\noindent
where $\lambda^{-2}$ measures the instanton size.

This choice automatically determines (up to a proportionality
constant) the electric part of the non-Abelian vectors for solution
D2. For the solution D1, the harmonic function $Q$ still has to be
specified, but if it is only a function of the radial coordinate, it
has to be proportional to $P$

\begin{equation}
Q=\xi_{1}P\, ,
\end{equation}

\noindent
for some constant $\xi_{1}$.\footnote{The additive constant in $Q$ is
  irrelevent.}

Hence, for the two classes of solutions D1 and D2, we have

\begin{equation}
  Z_{A}
  =
  \,\,\,\,
  \left\{
    \begin{array}{lr}
      {\displaystyle
      -\delta_{A\, i}\,\frac{\xi_{1}}{g\left(1+\lambda^{2} r\right)r}\frac{x^{i}}{r}\, ,
      }
      \hspace{1cm}
      &
        (\text{D1})
      \\
      \\
      {\displaystyle
      \delta_{A\, 3}\,\frac{\lambda^{2}\xi_{2} r}{g\left(1+\lambda^{2} r\right)}\, .
      }
      \hspace{1cm}
      &
        (\text{D2})
    \end{array}
    \right.
\end{equation}

For the harmonic functions $Z_{0,+,-}^{(0)}$, we take

\begin{equation}
  \label{eq:harmonicfunctions}
Z^{(0)}_{0,+,-}=a_{0,+,-}+\frac{b_{0,+,-}}{r}\, ,
\end{equation}

\noindent
and the complete functions $Z_{0}, \tilde{Z}_{+}$ and $Z_{-}$
appearing in the metric Eq.~(\ref{eq:gen-metric}) read

\begin{eqnarray}
  \label{eq:Z0rotatingBHs}
  Z_{0}
  & = &
        a_{0}+\frac{b_{0}}{r}
        -\frac{1}{18g^{2}}\frac{1}{r \left(a_{H} r+b_{H}\right)
        \left(1+\lambda^{2} r\right)^{2}} \, ,
  \\
  \nonumber\\
  \label{eq:Z-}
  Z_{-}
  & = &
        a_{-}+\frac{b_{-}}{r}\, , 
  \nonumber\\
  \label{eq:Z+}
  \tilde{Z}_{+} & = &
                  \left\{
                  \begin{array}{lr}
                    % Z^{(0)}_{+}
                    % -\frac{\xi_{1}^{2} \partial_{\underline{i}} P \partial_{\underline{i}} P} {g^{2} P^{2} Z_{-}}
                    % =
                    {\displaystyle
                    a_{+}+\frac{b_{+}}{r}
                    -\frac{\xi_{1}^{2}}{g^{2} r \left(a_{-} r+b_{-}\right)\left(1+\lambda^{2} r\right)^{2}}\, ,
                    }
                    \hspace{1cm}
                    & (\text{D1})
                    \\                  
                    \\
                    % Z_{+}^{(0)}-\frac{\xi_{2}^{2}}{g^{2} P^{2} Z_{-}}
                    % =
                    {\displaystyle
                    a_{+}+\frac{b_{+}}{r}
                    -\frac{\xi_{2}^{2} \lambda^{4} r^{3}}{g^{2} \left(a_{-} r+b_{-}\right)\left(1+\lambda^{2} r\right)^{2}}\, .
                    }
                    \hspace{1cm}
                    &(\text{D2})
                  \end{array}
                  \right.
\end{eqnarray}

Since we have restricted ourselves to GH spaces, the 1-form $\omega$
takes the form Eq.~(\ref{eq:sol2omega}) with $\omega_{5}$ given by
Eq.~(\ref{eq:omega5}) and with $\breve{\omega}$ implicitly determined by
Eq.~(\ref{eq:generalsolomega}). Choosing the harmonic function $M$ as

\begin{equation}
M =a_{M}+\frac{b_{M}}{r}\, ,  
\end{equation}

\noindent
one finds 

\begin{eqnarray}
  \label{eq:particularsolomega1}
  \omega_{5}
&  = &
  \left\{
    \begin{array}{lr}
      {\displaystyle
      a_{M}+\frac{b_{M}}{r}+\frac{\sqrt{3} \xi_{1}}{2g^{2}} \frac{1}{r\left(a_{H} r+b_{H}\right)\left(1+\lambda^{2} r\right)^{2}}\, ,
      }
      \hspace{1cm}
      &
        (\text{D1})
      \\
      \\
      {\displaystyle
      a_{M}+\frac{b_{M}}{r}-\frac{\sqrt{3}\xi_{2}}{2g^{2}}\frac{\lambda^{2} r \cos{\theta}}{\left(a_{H}r+b_{H}\right)\left(1+\lambda^{2} r\right)^{2}}\, ,
      }
      \hspace{1cm}
      &
        (\text{D2})
    \end{array}
    \right.
  \\
  \nonumber \\
  \nonumber \\
  \label{eq:particularsolomega2}
  \breve{\omega}
& = &
  \left\{
    \begin{array}{lr}
      {\displaystyle
      \left( a_{H}b_{M}-a_{M} b_{H}\right)\cos{\theta} d\phi \, ,
      }
      \hspace{1cm}
      &
        (\text{D1})
      \\
      \\
      {\displaystyle
      \left[\left( a_{H}b_{M}-a_{M} b_{H}\right)\cos{\theta}-\frac{\sqrt{3}\xi_{2}}{2g^{2}}\frac{\lambda^{2}r\sin^{2}{\theta}}{\left(1+\lambda^{2}r\right)^{2}}\right]d\phi\, .
      }
      \hspace{1cm}
      &
        (\text{D2})
    \end{array}
    \right.
\end{eqnarray}

At this point, the solutions are fully specified, up to the choice of
integration constants. This choice is constrained by requirements of
regularity, asymptotic flatness etc., which demand a closer, case by
case, study.

In particular, as we have already mentioned, the $a_{H}=0$ and
$a_{H}\neq 0$ cases correspond to asymptotically-flat 5 and
4-dimensional solutions, respectively. It is natural to analyze them
separately.
 
%%%%%%%%%%%%%%%%%%%%%%%%%%%%%%%%%%%%%%%%%%%%%%%%%%%%%%%%%%%%%%%%%%%%%%
%%%%%%%%%%%%%%%%%%%%%%%%%%%%%%%%%%%%%%%%%%%%%%%%%%%%%%%%%%%%%%%%%%%%%%
%%%%%%%%%%%%%%%%%%%%%%%%%%%%%%%%%%%%%%%%%%%%%%%%%%%%%%%%%%%%%%%%%%%%%%
%%%%%%%%%%%%%%%%%%%%%%%%%%%%%%%%%%%%%%%%%%%%%%%%%%%%%%%%%%%%%%%%%%%%%%
\subsection{5-Dimensional black holes ($a_{H}=0$)}
%%%%%%%%%%%%%%%%%%%%%%%%%%%%%%%%%%%%%%%%%%%%%%%%%%%%%%%%%%%%%%%%%%%%%%
%%%%%%%%%%%%%%%%%%%%%%%%%%%%%%%%%%%%%%%%%%%%%%%%%%%%%%%%%%%%%%%%%%%%%%
%%%%%%%%%%%%%%%%%%%%%%%%%%%%%%%%%%%%%%%%%%%%%%%%%%%%%%%%%%%%%%%%%%%%%%
%%%%%%%%%%%%%%%%%%%%%%%%%%%%%%%%%%%%%%%%%%%%%%%%%%%%%%%%%%%%%%%%%%%%%%

When $a_{H}=0$, the change of variables $\rho^{2}=4b_{H}r$ brings
the metric Eq.~(\ref{eq:sphericallysymmetricGHmetric}) to the form

\begin{equation}
  d\sigma^{2}
  =
  d\rho^{2}+\frac{\rho^{2}}{4}
  \left(d\Psi^{2}+d\phi^{2}+d\theta^{2}+d\phi^{2}+2\cos{\theta} d\Psi d\phi\right)
  \equiv
  d\rho^{2}+\rho^{2}\, d\Omega^{2}_{(3)/n} \, ,
\end{equation}

\noindent
where we have introduced the angular coordinate
$\Psi=2\eta/(n\ell_{s})$ whose period is $\Psi\sim \Psi+4\pi/n$ and
where $d\Omega^{2}_{(3)/n}$ is the metric of the lens space
$S^{3}/\mathbb{Z}_{n}$. From now on we discuss the $n=1$ case, for
which the above metric is that of $\mathbb{E}^{4}$ and the
5-dimensional spacetime metric of the solution
Eq.~(\ref{eq:gen-metric}) can be cast in the form 

\begin{equation}
  \label{eq:5dmetric}
  ds^{2}
  =
  \left(\mathcal{Z}_{0}\tilde{\mathcal{Z}}_{+}\mathcal{Z}_{-}\right)^{-2/3}(dt+\omega)^{2}
  -\left(\mathcal{Z}_{0}\tilde{\mathcal{Z}}_{+}\mathcal{Z}_{-}\right)^{1/3}
  \left(d\rho ^{2}+\rho^{2}d\Omega^{2}_{(3)}\right)\, ,
\end{equation}

\noindent
where we have defined

\begin{equation}
  \mathcal{Z}_{0} \equiv Z_{0}/a_{0}\, ,
  \hspace{1cm}
  \mathcal{Z}_{-} \equiv Z_{-}/a_{-}\, ,
  \hspace{1cm}
  \tilde{\mathcal{Z}}_{+} \equiv \tilde{Z}_{+}/\tilde{a}_{+}\, ,
\end{equation}

\noindent
$\tilde{a}_{+}$ being the asymptotic value of $\tilde{Z}_{+}$, which
is given by

\begin{equation}
  \label{eq:asymptoticvalueZ+}
  \tilde{a}_{+}
  =
  \left\{
    \begin{array}{lr}
      a_{+}\, ,
      &
        \hspace{1cm}
        (\text{D1})
      \\
      & \\
      {\displaystyle
      a_{+}-\frac{\xi_{2}^{2}}{a_{-} g^{2}}\, .
      }
      &
        \hspace{1cm}
        (\text{D2})
    \end{array}
    \right.
\end{equation}

\noindent
and where we have imposed an asymptotic flatness condition on
$a_{0}, a_{-}$ and $\tilde{a}_{+}$, namely

\begin{equation}
  \label{eq:asyflatcond}
\frac{27}{2}a_{0} \tilde{a}_{+} a_{-}=1\, .
\end{equation}

This condition, together with the expressions of the scalars in terms
of the $Z$ functions Eq.~(\ref{eq:physcalarsp2}) allows us to write
$a_{0}, a_{-}$ and $\tilde{a}_{+}$ in terms of the two moduli of this
theory $\phi_{\infty},k_{\infty}$:

\begin{equation}
  a_{0}=\frac{1}{3}e^{\phi_{\infty}}k_{\infty}^{-2/3}\, ,
  \hspace{1cm}
  \tilde{a}_{+}=\frac{1}{3}k_{\infty}^{4/3}\, ,
  \hspace{1cm}
  a_{-}=\frac{2}{3}e^{-\phi_{\infty}}k_{\infty}^{-2/3} \, .
\end{equation}

We can, therefore, eliminate these three integration constants in the
functions that appear in the metric, which, with the definitions

\begin{equation}
  \kappa^{2}\equiv 4b_{H}\lambda^{-2}\, ,
  \hspace{1cm}
\tilde \xi_{1}\equiv 4b_{H} \xi_{1}\, ,  
\end{equation}

\noindent
now take the form

\begin{eqnarray}
  \mathcal{Z}_{0}
  & = &
        1+ \frac{\mathcal{Q}_{0}}{\rho^{2}}
        +\frac{2\, e^{-\phi_{\infty}}k_{\infty}^{2/3}}{3 g^{2}}
        \frac{\rho^{2}+2\kappa^{2}}{(\rho^{2}+\kappa^{2})^{2}}\, ,
  \\
\nonumber\\
  \mathcal{Z}_{-}
  & = &
        1+\frac{\mathcal{Q}_{-}}{\rho^{2}}\, ,
                      \\
  \nonumber\\
  \tilde{\mathcal{Z}}_{+}
  & = &
        \left\{
        \begin{array}{lr}
          {\displaystyle
          1+\frac{\tilde{\mathcal{Q}}_{+}}{\rho^{2}}
          +\frac{9 \tilde \xi_{1}^{2}\, e^{\phi_{\infty}}k_{\infty}^{-2/3} }{ 2g^{2}}
          \frac{\rho^{4}+\rho^{2}\left(\mathcal{Q}_{-}+2\kappa^{2}\right)
          +\kappa^{2}\left(\kappa^{2}+2\mathcal{Q}_{-}\right)}{\mathcal{Q}_{-}
          \left(\rho^{2}+\mathcal{Q}_{-}\right)\left(\rho^{2}+\kappa^{2}\right)^{2}}\, ,
          }
          &
            \hspace{.3cm}
            \text{(D1)}
          \\
          \\
          {\displaystyle
          1+\frac{\tilde{\mathcal{Q}}_{+}}{\rho^{2}}
          +\frac{9 \xi_{2}^{2}\, e^{\phi_{\infty}}k_{\infty}^{-2/3} }{ 2g^{2}}
          \frac{\rho^{4}\left(\mathcal{Q}_{-}+2\kappa^{2}\right)+\rho^{2}\kappa^{2}
          \left(2\mathcal{Q}_{-}+\kappa^{2}\right)+\mathcal{Q}_{-} \kappa^{4}}{
          \left(\rho^{2}+\mathcal{Q}_{-}\right)\left(\rho^{2}+\kappa^{2}\right)^{2}}\, .
          }
          &
            \hspace{.3cm}
            \text{(D2)}
        \end{array}
        \right.
\end{eqnarray} 

\noindent
Here we have introduced new constants
$\mathcal{Q}_{0}, \tilde{\mathcal{Q}}_{+}$ and $\mathcal{Q}_{-}$ whose
relation with the parameters of the harmonic functions
$b_{0},b_{+},b_{-},b_{H}$ in Eqs.~(\ref{eq:harmonicfunctions}) is

\begin{equation}
  \mathcal{Q}_{0}=\frac{4b_{H}}{a_{0}}\left(b_{0}-\frac{1}{18 {g}^{2}}\right)\, ,
  \hspace{.5cm}
  \mathcal{Q}_{-}=\frac{4 b_{H}b_{-}}{a_{-}}\, ,
  \hspace{.5cm}
  \tilde{\mathcal{Q}}_{+}
  =
  \left\{
    \begin{array}{lr}
      {\displaystyle
      \frac{4b_{H}}{\tilde{a}_{+}}\left(b_{+}-\frac{\xi_{1}^{2}}{g^{2}b_{-}}\right)\, ,
      }
      &
        \hspace{.1cm}
        \text{(D1)}
      \\
      \\
      {\displaystyle
      \frac{4b_{H} b_{+}}{\tilde{a}_{+}}\, ,
      }
      &
        \hspace{.1cm}
        \text{(D2)}        
    \end{array}
    \right.
\end{equation}

Finally, setting $a_{M}=0$ in Eqs.~(\ref{eq:particularsolomega1}) and
(\ref{eq:particularsolomega2}), we find

\begin{equation}
  \omega
  =
  \left\{
    \begin{array}{lr}
      {\displaystyle
      \left[
      \frac{\mathcal{J}+\frac{\sqrt{3}\tilde \xi_{1}}{2g^{2}}}{\rho^{2}}
      -\frac{\sqrt{3}\tilde \xi_{1}}{2g^{2}}
      \frac{\rho^{2}+2\kappa^{2}}{\left(\rho^{2}+\kappa^{2}\right)^{2}}
      \right]
      \left(d\Psi+\cos{\theta} d\phi\right)\, ,
      }
      &
          \hspace{.1cm}
        \text{(D1)}
      \\
      \\
      {\displaystyle
      \frac{\mathcal{J}}{\rho^{2}}\left(d\Psi+\cos{\theta} d\phi\right)
      -\frac{\sqrt{3}\,  \xi_{2}\, \kappa^{2}\rho^{2}}{2g^{2}(\rho^{2}+\kappa^{2})^{2}}
      \left(d\phi+\cos{\theta} d\Psi\right)\, ,
      }
      &
          \hspace{.1cm}
        \text{(D2)}
    \end{array}
    \right.
\end{equation}
where we have defined a new constant

\begin{equation}
\mathcal{J}=4b_{M} b_{H}^{2}\, .
\end{equation}

We have replaced some integration constants by physical quantities and
we have also made some redefinitions. Then, at this point, the
solutions we have constructed depend on the independent constants 

\begin{displaymath}
  \phi_{\infty}\, ,\,\,\,
  k_{\infty}\, ,\,\,\,
  g\, ,\,\,\,
  \kappa\, ,\,\,\,
  \mathcal{J}\, ,\,\,\,
  \mathcal{Q}_{0}\, ,\,\,\,
  \tilde{\mathcal{Q}}_{+}\, ,\,\,\,
  \mathcal{Q}_{-}\, ,\,\,\,\text{and}\,\,\,
  \tilde{\xi}_{1}\,\,\, \text{or}\,\,\,\xi_{2}\, .
\end{displaymath}

\noindent
The first three of these have a clear physical meaning: they are
moduli of the solutions: asymptotic values of two scalars and
Yang-Mills coupling constant. There is another modulus of the
solutions: the asymptotic value of the gauge-invariant combination
$\sqrt{\ell_{A}\ell_{A}}$, that we can denote by $v$. Only
for the D2 solution it has a non-trivial value:

\begin{equation}
 v
  =
  \frac{\xi_{2}}{a_{-}g}
  =
  \frac{3}{2g}\xi_{2} e^{\phi_{\infty}} k_{\infty}^{2/3}\, , \hspace{1.5cm}\text{(D2)}
\end{equation}

\noindent
which allows us to replace $\xi_{2}$ by
$\frac{2}{3}g v e^{-\phi_{\infty}} k_{\infty}^{-2/3}$.

Our next task will be to compute the
conserved charges of the solution in terms of the rest of the
integration constants.

%%%%%%%%%%%%%%%%%%%%%%%%%%%%%%%%%%%%%%%%%%%%%%%%%%%%%%%%%%%%%%%%%%%%%% 
%%%%%%%%%%%%%%%%%%%%%%%%%%%%%%%%%%%%%%%%%%%%%%%%%%%%%%%%%%%%%%%%%%%%%%
%%%%%%%%%%%%%%%%%%%%%%%%%%%%%%%%%%%%%%%%%%%%%%%%%%%%%%%%%%%%%%%%%%%%%%
%%%%%%%%%%%%%%%%%%%%%%%%%%%%%%%%%%%%%%%%%%%%%%%%%%%%%%%%%%%%%%%%%%%%%%
\subsubsection*{Charges of the solution}
%%%%%%%%%%%%%%%%%%%%%%%%%%%%%%%%%%%%%%%%%%%%%%%%%%%%%%%%%%%%%%%%%%%%%%
%%%%%%%%%%%%%%%%%%%%%%%%%%%%%%%%%%%%%%%%%%%%%%%%%%%%%%%%%%%%%%%%%%%%%%
%%%%%%%%%%%%%%%%%%%%%%%%%%%%%%%%%%%%%%%%%%%%%%%%%%%%%%%%%%%%%%%%%%%%%%
%%%%%%%%%%%%%%%%%%%%%%%%%%%%%%%%%%%%%%%%%%%%%%%%%%%%%%%%%%%%%%%%%%%%%%

It is well-known that the presence of Chern-Simons terms in field
strengths or actions leads to the possibility of defining different
notions of charge, see for instance
Refs.~\cite{Townsend:1996em,Marolf:2000cb}. In the theories under
consideration, they induce the occurrence of $F\wedge F$ terms in the
equations of motion of the vector fields, as can be seen in
Eq.~(\ref{eq:vectoreom}), which in differential-form language takes
the form:

\begin{equation}
  \label{eq:vectoreomdiff}
  -\mathfrak{D}(a_{IJ}\star F^{J})
  +\tfrac{1}{\sqrt{3}}C_{IJK}F^{J}\wedge F^{K}
  +gk_{I\, x}\mathfrak{D}\phi^{x}
  =
  0\, .
\end{equation}

\noindent
The $F\wedge F$ terms vanish when all the vector fields are purely
electric and static, but they give non-vanishing contributions at
infinity when they are non-static or magnetic (instantonic, for
instance). Therefore, one gets different results in the calculation of
a charge, depending on whether one includes these terms in the
definition or not.

Let us study the different possibilities.

If we couple the supergravity action to a $0$-brane that couples
electrically to the vector field $A^{I}$, the equations of motion
Eq.~(\ref{eq:vectoreomdiff}) are modified by a 1-form current
$J^{S}_{I}$ as follows\footnote{Apart from the overall factor of
  $16\pi G_{N}^{(5)}$, that we are ignoring here, typically the 1-form
  current $J^{S}_{I}$ will come multiplied by different combinations
  of asymptotic vallues of the scalars and other constants that we
  will ignore in our discussion. }

\begin{equation}
  \label{eq:vectoreomdiffsource}
 - \mathfrak{D}(a_{IJ}\star F^{J})
  +\tfrac{1}{\sqrt{3}}C_{IJK}F^{J}\wedge F^{K}
  +gk_{I\, x}\mathfrak{D}\phi^{x}
  =
  \star J_{I}{}^{S}\, ,
\end{equation}

\noindent
and we can compute the so-called \textit{brane-source charges},
$\mathcal{Q}_{I}^{S}$, by integrating both sides over some spatial
4-volume (such as a $t=\mathrm{constant}$ hypersurface):

\begin{equation}
  \mathcal{Q}_{I}^{S}
\equiv
  \int_{V^{4}}  \star J_{I}{}^{S}
  =
  \int_{V^{4}}
  \left\{
     - \mathfrak{D}(a_{IJ}\star F^{J})
  +\tfrac{1}{\sqrt{3}}C_{IJK}F^{J}\wedge F^{K}
  +gk_{I\, x}\mathfrak{D}\phi^{x}
    \right\}\, .
\end{equation}

In general, this charge is not conserved, $d\star J_{I}{}^{S}\neq 0$,
because the l.h.s~of Eq.~(\ref{eq:vectoreomdiffsource}) is not
closed. However, in the ungauged directions, the Killing vectors
$k_{I}{}^{x}$ vanish, the gauge-covariant derivative becomes an
ordinary exterior derivative and the $F\wedge F$ terms are a closed
(but not exact) 4-form and

\begin{equation}
  \mathcal{Q}_{I}^{S}
  =
  \frac{1}{2\pi^{2}}
  \int_{V^{4}}
  \left\{
     -d(a_{IJ}\star F^{J})
  +\tfrac{1}{\sqrt{3}}C_{IJK}F^{J}\wedge F^{K}
    \right\}\, ,
\end{equation}

\noindent
is a conserved charge.
%\begin{equation}
%  \omega^{\rm YM}
%  =
%  A^{A}\wedge dA^{A}+\tfrac{1}{3}\epsilon^{ABC}{A}^{A}\wedge{A}^{B}\wedge{A}^{C}\, ,
%  \hspace{1cm}
%  d\omega^{\rm YM} = F^{A}\wedge F^{A}\, .
%\end{equation}

Since each of the two terms that appear in the above integral for the
Abelian directions are closed 4-forms, we can use them separately to
define other possible conserved charges. In particular, using only the
terms with second derivatives in the volume integral gives the
so-called \textit{Maxwell charges}

\begin{equation}
  \mathcal{Q}_{I}^{M}
\equiv
-  \frac{1}{2\pi^{2}}\int_{V^{4}} d(a_{IJ}\star F^{J})\, .
\end{equation}

For our supergravity model (see the action
Eq.~(\ref{eq:gaugedST[2,6]action})) and applying Stokes theorem, we
have
 
\begin{eqnarray}
  \mathcal{Q}_{0}^{M}
  & = &
        -\frac{1}{6\pi^{2}}\int_{\partial V_{4}}e^{2\phi}k^{-4/3}\star F^{0}\, ,
  \\
  \nonumber \\
  \mathcal {Q}_{+}^{M}
  & = &
        -\frac{1}{6\pi^{2}}\int_{\partial V_{4}}
        \left\{
        \frac{1}{4}k^{8/3}\left(1+2e^{-\phi}k^{-2}\ell^{B}\ell^{B}\right)^{2}\star F^{+}
        + e^{-2\phi}k^{-4/3}\ell^{B}\ell^{B}\star F^{-}
        \right.
        \nonumber\\
  \nonumber \\
  &&
     \left.
     +e^{-\phi}k^{2/3}\left(1+2e^{-\phi}k^{-2}\ell^{B}\ell^{B}\right)\ell^{A}\star F^{A}
     \right\}\, ,
  \\
  \nonumber\\
  \mathcal{Q}_{-}^{M}
  & = &
        -\frac{1}{6\pi^{2}}\int_{\partial V_{4}}e^{-2\phi}k^{-4/3}
        \star\left(F^{-}+\ell^{B}\ell^{B} F^{+}+2\ell^{A} F^{A}\right)\, ,
\end{eqnarray}

\noindent
where $\partial V_{4}$ is the boundary of $V_{4}$. The relation
between the Maxwell charges and the brane-source charges
is\footnote{Observe that, in general, we cannot apply Stokes' theorem
  since they are closed but not exact 4-forms.  }

\begin{eqnarray}
  \mathcal{Q}_{0}^{S}
  & = &
        \mathcal{Q}_{0}^{M}+\frac{1}{12\sqrt{3}\pi^{2}}
        \int_{V_{4}}\left(F^{+}\wedge F^{-}-F^{A}\wedge F^{A}\right)\, ,
  \\
\nonumber\\
  \mathcal {Q}_{\pm}^{S}
  & = &
        \mathcal{Q}_{\pm}^{M}+\frac{1}{6\sqrt{3}\pi^{2}}\int_{V_{4}}F^{0}\wedge F^{\mp}\, .
\end{eqnarray}

By direct computation, we find that, for the solutions described in
this section, the Maxwell charges have the following values

\begin{eqnarray}
  \mathcal{Q}_{0}^{M}
  & = &
        \frac{2}{\sqrt{3}}e^{\phi_{\infty}}k_{\infty}^{-2/3}\mathcal{Q}_{0}^{\infty}\, ,
  \\
\nonumber\\
  \mathcal{Q}_{+}^{M}
  & = &
        \left\{
        \begin{array}{lr}
          \displaystyle{\frac{1}{\sqrt{3}}
          k_{\infty}^{4/3}\tilde{\mathcal{Q}}^{\infty}_{+}}
          &
            \hspace{.1cm} \text{(D1)}
          \\
          \\
          \displaystyle{\frac{1}{\sqrt{3}}k_{\infty}^{4/3}\tilde{\mathcal{Q}}_{+}}
          &
            \hspace{.1cm}\text{(D2)}
          \\
        \end{array}
  \right.\,,
\\
\nonumber\\
  \mathcal{Q}_{-}^{M}
  & = &
        \left\{
        \begin{array}{lr}
          \displaystyle{\frac{2}{\sqrt{3}}
          e^{-\phi_{\infty}}k_{\infty}^{-2/3}\mathcal{Q}^{\infty}_{-}}
          &
            \hspace{.1cm} \text{(D1)}
          \\
          \\
          \displaystyle{\frac{2}{\sqrt{3}}
          e^{-\phi_{\infty}}k_{\infty}^{-2/3}
          \left(
          1+2v^2e^{-\phi_{\infty}}k_{\infty}^{-2}
          \right)\mathcal{Q}^{\infty}_{-}}
          &
            \hspace{.1cm}\text{(D2)}
          \\
        \end{array}
  \right.\,,
\end{eqnarray}

\noindent
where we have defined
$\mathcal{Q}_{0}^{\infty}, \tilde{\mathcal{Q}}_{+}^{\infty}$ and
$\mathcal{Q}^{\infty}_{-}$ as

\begin{eqnarray}
  \mathcal{Q}^{\infty}_{0}
  & \equiv &
             \lim_{\rho \rightarrow \infty} \rho^{2}
             \left({\mathcal Z}_{0}-1\right)
             =
             \mathcal{Q}_{0}+\frac{2e^{-\phi_{\infty}}k_{\infty}^{2/3}}{3g^{2}}\, ,
  \\
\nonumber\\
  \mathcal {\tilde Q}^{\infty}_{+}
  & \equiv &
             \lim_{\rho \rightarrow \infty} \rho^{2}
             \left(\mathcal {\tilde{Z}}_{+}-1\right)
             =
             \left\{
        \begin{array}{lr}
          {\displaystyle
          \tilde{\mathcal{Q}}_{+}\left(
          1+\frac{9\tilde \xi_{1}^{2}e^{\phi_{\infty}}k_{\infty}^{-2/3}}{2g^{2}\tilde{\mathcal{Q}}_{+}\mathcal{Q}_{-}}\right) 
          }
          &
            \hspace{.1cm} \text{(D1)}
          \\
          \\
          {\displaystyle   \tilde{\mathcal{Q}}_{+}
          +2v^2e^{-\phi_{\infty}}k_{\infty}^{-2}
          \left(\mathcal{Q}_{-}+2\kappa^{2}\right)
          }
          &
            \hspace{.1cm}
            \text{(D2)}
        \end{array}
            \right.\,, \\
  \nonumber
  \mathcal{Q}^{\infty}_{-}
  & \equiv &
             \lim_{\rho \rightarrow \infty} \rho^{2}
             \left({\mathcal Z}_{-}-1\right)=\mathcal{Q}_{-}\, .
\end{eqnarray}

Observe that the difference between $\mathcal{Q}^{\infty}_i$ and the
constants $\mathcal{Q}_i$ is always a shift by some quantity. This behavior
is characteristic of systems which have delocalized sources (such as those
introduced by the non-Abelian fields) that can contribute at
infinity. %In some cases, the contribution is
%such that near-horizon charges get screened.\footnote{This is
%precisely what happens with heterotic black holes
%when higher-curvature terms ($\alpha'$-corrections) are taken into account 
%\cite{Cano:2018qev, Cano:2018brq}, being particularly relevant for the study of the 
%so-called small black holes \cite{Cano:2018hut}.} 
In this respect, the shift in $\mathcal{Q}_{0}$ corresponds to the contribution of 
the instanton to this kind of charge, already observed in
Refs.~\cite{Cano:2017qrq, Cano:2018qev,Cano:2018brq, Cano:2017sqy}. 
The new shifts in $\mathcal{\tilde Q}_{+}$, which are proportional respectively to
$\tilde{\xi}_{1}^{2}$ and $v^2$ (or equivalently to $\xi_{2}^{2}$), are due to the
 dyonic nature of the instanton.

The integrals of the $F\wedge F$ terms are 

\begin{eqnarray}
\label{eq:F0F+}
\frac{1}{2\pi^{2}}  \int_{V_{4}}F^{0}\wedge F^{+}
  & = &
        12 e^{-\phi_{\infty}}k_{\infty}^{-2/3} \,
        \frac{\beta}{1+\beta}\mathcal{Q}_{-}\, ,
  \\
\nonumber\\
\nonumber\\
\label{eq:F0F-}
\frac{1}{2\pi^{2}} \int_{V_{4}}F^{0}\wedge F^{-}
  & = &
        \left\{
        \begin{array}{lr}
          \displaystyle{
          6{k}_{\infty}^{4/3}
          \,\frac{\beta}{1+\beta}\tilde{\mathcal{Q}}^{\infty}_{+}\, ,
          }
          &
            \hspace{1cm}\text{(D1)}
          \\
\\
          \displaystyle{
          6{k}_{\infty}^{4/3} \,
          \frac{\beta}{1+\beta}\tilde{\mathcal{Q}}_{+}\, ,
          }
          &
            \hspace{1cm}\text{(D2)}
          \\
        \end{array}
  \right. 
  \\
  \nonumber\\
  \nonumber\\
\label{eq:F+F-}
\frac{1}{2\pi^{2}}  \int_{V_{4}}F^{+}\wedge F^{-}
  & = &
        \left\{
        \begin{array}{lr}
          \displaystyle{
          12 e^{\phi_{\infty}}{k}_{\infty}^{-2/3} \, 
          \frac{\beta}{1+\beta}}
          \left(
          1+\frac{9 e^{\phi_{\infty}}k_{\infty}^{-2/3}\tilde{\xi}_{1}^{2}}{2g^{2}
          \tilde{\mathcal{Q}}_{+} \mathcal{Q}_{-}}
          \right)\mathcal{Q}_{0}\, ,
          &
            \hspace{0.1cm}\text{(D1)}\\
          \\
          \displaystyle{12e^{\phi_{\infty}}{k}_{\infty}^{-2/3}  \,
          \frac{\beta}{1+\beta}}\mathcal{Q}_{0}\, ,
          &
            \hspace{0.1cm}\text{(D2)}
          \\
        \end{array}
  \right.
  \\
\nonumber\\
\nonumber\\
\label{eq:instantonnumber}
  \frac{1}{2\pi^{2}}  \int_{V_{4}}F^{A}\wedge F^{A}
  & = &
        \left\{
        \begin{array}{cc}
          \displaystyle{\frac{8}{g^{2}}+ 54
          e^{2\phi_\infty}k_{\infty}^{-4/3}\mathcal Q_0
          \frac{\beta}{1+\beta}\frac{\tilde \xi_1^2}{g^2\tilde{\mathcal
          Q_+}\mathcal Q_-}}\, ,
          &
            \hspace{0.1cm}\text{(D1)}\\
          \\
          \displaystyle{\frac{8}{g^{2}}}\, ,
           &
          \hspace{0.1cm}\text{(D2)}
  \end{array}\right.
\end{eqnarray}

\noindent
where the constant $\beta$ is defined by 

\begin{equation}
  \label{eq:betadef}
  \beta
  =\,
  \frac{4\tilde {\mathcal{J}}^{2}} {\mathcal{Q}_{0} \tilde{\mathcal{Q}}_{+}\mathcal{Q}_{-}
    -4\tilde {\mathcal{J}^{2}}}\, ,
  \quad
  \text{with}
  \quad
  \tilde{\mathcal{J}}
  =
  \left\{
    \begin{array}{lr}
      \displaystyle{\mathcal{J}+\frac{\sqrt{3}\tilde{\xi}_{1}}{2g^{2}}}\, ,
      &
        \hspace{0.1cm}\text{(D1)}
      \\
\\
      \displaystyle{\mathcal{J}}\, ,
      &
        \hspace{0.1cm}\text{(D2)}
      \\
    \end{array}
  \right.
\end{equation}

\noindent
and $\mathcal{J}$ is related to the angular momenta of the solutions, 
as we will see below. The shift in $\tilde{\mathcal{J}}$ is due to
the contribution of the non-Abelian field to the angular momentum at
the horizon.

%Eq.~(\ref{eq:instantonnumber}) indicates that our solutions include a
%BPST instanton \cite{Belavin:1975fg}, or, more precisely, a dyonic deformation of it. The
%electric part of this dyonic configuration, mentioned above, will be
%characterized later on.
Eq.~(\ref{eq:instantonnumber}) indicates that our solutions include a dyonic 
deformation of the BPST instanton \cite{Belavin:1975fg}. The electric part of this
dyonic configuration will be characterized later on. Notice that
the instanton number in the D1 case is not quantized since the integral $\int_{V_4} F^A\wedge F^A$ 
has a second contribution due to the fact that the gauge fields do not vanish at the horizon.
 
Taking all these results into account, we find that the brane-source
charges are given by

\begin{eqnarray}
  \mathcal{Q}_{0}^{S}
  & = &
     \frac{2}{\sqrt{3}}e^{\phi_{\infty}}k_{\infty}^{-2/3}
          \left(1+\frac{\beta}{1+\beta}
          \right){\mathcal Q}_0\, , \\
  \nonumber\\
  \mathcal{Q}_{+}^{S}
  & = &
        \left\{
        \begin{array}{lr}
          \displaystyle{
          \frac{1}{\sqrt{3}}k_{\infty}^{4/3}}
          \left(1+\frac{\beta}{1+\beta}\right)\tilde{\mathcal{Q}}^{\infty}_{+}\, ,
          &
            \hspace{1cm} \text{(D1)}
          \\
          \\
          \displaystyle{\frac{1}{\sqrt{3}}k_{\infty}^{4/3}}
          \left(1+\frac{\beta}{1+\beta}\right)\tilde{\mathcal{Q}}_{+}\, ,
          &
            \hspace{1cm}\text{(D2)}
          \\
        \end{array}
  \right.
  \\
  \nonumber\\
    \nonumber\\
  \mathcal{Q}_{-}^{S}
  & = &
        \left\{
        \begin{array}{lr}
          \displaystyle{
          \frac{2}{\sqrt{3}}e^{-\phi_{\infty}}k_{\infty}^{-2/3}}
          \left(1+\frac{\beta}{1+\beta}\right)\mathcal{Q}_{-}\, ,
          &
            \hspace{.1cm} \text{(D1)}
          \\
          \\
          \displaystyle{
          \frac{2}{\sqrt{3}}e^{-\phi_{\infty}}k_{\infty}^{-2/3}}
          \left(1+2v^2e^{-\phi_{\infty}}k_{\infty}^{-2}
          +\frac{\beta}{1+\beta}\right)\mathcal{Q}_{-}\, .
          &
            \hspace{.1cm}\text{(D2)}
          \\
        \end{array}
  \right.
\end{eqnarray}

In order to characterize the electric part of the non-Abelian dyonic
configuration, we can integrate the gauge-invariant quantity
$\ell^{A}\star F^{A}$ over a $S^{3}$ and take the $\rho\to \infty $ limit or
that in which it goes to zero. % to capture delocalized electric-type charges
%that can be seen at infinity or localized electric-type charges which are
%screened at infinity, respectively.  The D1 and D2 cases differ, precisely, in
%the delocalized versus localized natures of these charges.  
For the solution D1, the profile of the fields is such that $\int_{S^3}\ell^A\star F^A$
vanishes when computed in the $\rho\to \infty $ limit  since they fall off to zero too fast, but in
the $\rho\to 0$ limit it does not. The opposite is true for the D2 solution.
Thus, we find\footnote{$\tilde{\xi}_{1}/g$ has dimensions of length squared,
  as a charge, while $\xi_{2}/g$ is dimensionless.}

\begin{eqnarray}
& &  \frac{1}{2\pi^{2}} \int_{S^{3}_{0}}\ell^{A}\star F^{A}
        =
        9\sqrt{3}\frac{\tilde{\xi}_{1}^{2}e^{2\phi_{\infty}}}{g^{2}}
        \left(\frac{\mathcal{Q}_{0}^{2}}{\tilde{\mathcal{Q}}_{+}\mathcal{Q}_{-}^{4}}
        \right)^{1/3}\, ,
        %\hspace{.1cm} \text{(D1)}
  \\
  \nonumber   \\
  \label{eq:chargeD2}
  \mathcal {Q}_{\rm{D2}}
  & = &
        \frac{1}{2\pi^{2}} \int_{S^{3}_{\infty}}\ell^{A}\star F^{A}
        =
        9\sqrt{3}\frac{\xi_{2}^{2}e^{2\phi_{\infty}}}{g^{2}}
        \left(\kappa^{2} +\tilde{\mathcal{Q}}^{\infty}_{+}+\mathcal{Q}^{\infty}_{-}
        \right)\, .
        %\hspace{.1cm}\text{(D2)}
\end{eqnarray}
It is worth mentioning that the while the interpretation of $\int_{S_0^3}\ell^A\star F^A$ as a charge is not very rigorous, the quantity that we have denoted by $\mathcal Q_{D_2}$ does have a charge interpretation. It is the  charge (up to moduli factors) associated of the unbroken $U(1)$ vector field \cite{Eyras:2000dg}.
%%%%%%%%%%%%%%%%%%%%%%%%%%%%%%%%%%%%%%%%%%%%%%%%%%%%%%%%%%%%%%%%%%%%%%
%%%%%%%%%%%%%%%%%%%%%%%%%%%%%%%%%%%%%%%%%%%%%%%%%%%%%%%%%%%%%%%%%%%%%%
%%%%%%%%%%%%%%%%%%%%%%%%%%%%%%%%%%%%%%%%%%%%%%%%%%%%%%%%%%%%%%%%%%%%%%
\subsubsection*{Mass and angular momenta}
%%%%%%%%%%%%%%%%%%%%%%%%%%%%%%%%%%%%%%%%%%%%%%%%%%%%%%%%%%%%%%%%%%%%%%
%%%%%%%%%%%%%%%%%%%%%%%%%%%%%%%%%%%%%%%%%%%%%%%%%%%%%%%%%%%%%%%%%%%%%%
%%%%%%%%%%%%%%%%%%%%%%%%%%%%%%%%%%%%%%%%%%%%%%%%%%%%%%%%%%%%%%%%%%%%%%

The mass and the two independent angular momenta of the solution can
be found by examining the asymptotic behavior of the metric
Eq.~(\ref{eq:5dmetric}) in a suitable coordinate system. Thus, to this
aim, it is convenient to introduce a new set of coordinates
$(\tilde{t}, \tilde{\rho}, \tilde{\theta}, \tilde{\phi}_{+},
\tilde{\phi}_{-})$ related to the previous one by the following
coordinate transformation

\begin{equation}
\begin{aligned}
\tilde{t}=\,t\, ,\quad \tilde{\rho}=\,\rho \left(\mathcal{Z}_{0}\tilde{\mathcal{Z}}_{+}\mathcal{Z}_{-}\right)^{1/3}\, ,\quad \tilde{\theta}=\,\frac{\theta}{2}\, \, ,\quad
\tilde{\phi}_{\pm}=\,\frac{\Psi\pm \phi}{2}\, .
\end{aligned}
\end{equation}

In terms of these new coordinates, the asymptotic expansion of
Eq.~(\ref{eq:5dmetric}) for large values of $\tilde{\rho}$ reads

\begin{equation}
\begin{aligned}
  ds^{2}
  & \sim
  \left(1-\frac{8G_{N}^{(5)} \mathcal{M}}{3\pi \tilde{\rho}^{2}}\right)d\tilde{t}^{2}
  +\frac{4\mathcal{J}_{+}}{\tilde{\rho}^{2}}
  \cos^{2}\tilde{\theta}\, d\tilde{t} \,d\tilde{\phi}_{+}
  +\frac{4\mathcal{J}_{-}}{\tilde\rho^{2}} \sin^{2}\tilde{\theta}\, dt\,d\tilde{\phi}_{-}
  \\
  \\
  &
  -\left(1+\frac{8G_{N} ^{(5)} \mathcal{M}}{3\pi\tilde{\rho}^{2}}\right)d\tilde{\rho}^{2}
  -\tilde{\rho}^{2}\left(d\tilde{\theta}^{2}+\cos^{2}\tilde{\theta}  \,d
    \tilde{\phi}_{+}^{2}+\sin^{2}\tilde{\theta} \,d\tilde{\phi}_{-}^{2}\right)\, ,
\end{aligned}
\end{equation}

\noindent
where $\mathcal{M}$, the ADM mass of the solution, is given by

\begin{equation}
  \label{eq:mass5dbhs}
  \mathcal{M}
  =
  \frac{\pi}{4 G_{N}^{(5)}}
  \left(\mathcal{Q}^{\infty}_{0}+\tilde{\mathcal{Q}}^{\infty}_{+}
    +\mathcal{Q}^{\infty}_{-}\right)\, ,
\end{equation}

\noindent
and $\mathcal{J}_{\pm}$ are the two independent angular momenta of the solution

\begin{equation}\label{eq:angularmomenta5d}
\mathcal J_{\pm}=\left\{\begin{array}{cc}
                          \mathcal J\, ,
                          &
                            \hspace{1.5cm} \text{(D1)}\\
\\
                          \displaystyle{\mathcal{J}\mp\frac{\sqrt{3}   \kappa^{2}\xi_{2}}{2g^{2}}}\, .
                          &
                            \hspace{1.5cm} \text{(D2)}
\end{array}\right.
\end{equation}

%%%%%%%%%%%%%%%%%%%%%%%%%%%%%%%%%%%%%%%%%%%%%%%%%%%%%%%%%%%%%%%%%%%%%%
%%%%%%%%%%%%%%%%%%%%%%%%%%%%%%%%%%%%%%%%%%%%%%%%%%%%%%%%%%%%%%%%%%%%%%
%%%%%%%%%%%%%%%%%%%%%%%%%%%%%%%%%%%%%%%%%%%%%%%%%%%%%%%%%%%%%%%%%%%%%%
\subsubsection*{Properties of the solution}
%%%%%%%%%%%%%%%%%%%%%%%%%%%%%%%%%%%%%%%%%%%%%%%%%%%%%%%%%%%%%%%%%%%%%%
%%%%%%%%%%%%%%%%%%%%%%%%%%%%%%%%%%%%%%%%%%%%%%%%%%%%%%%%%%%%%%%%%%%%%%
%%%%%%%%%%%%%%%%%%%%%%%%%%%%%%%%%%%%%%%%%%%%%%%%%%%%%%%%%%%%%%%%%%%%%%

Let us list here the main properties of these solutions:

\begin{itemize}

\item There is a regular horizon located at $\rho=0$. Hence, they
  describe supersymmetric, rotating, asymptotically-flat black
  holes.  The induced metric at the horizon is

\begin{equation}
  -ds_{\rm H}^{2}
  =
  \frac{\left(\mathcal{Q}_{0} \tilde{\mathcal{Q}}_{+}\mathcal{Q}_{-}\right)^{1/3}}{4}
  \left[\frac{1}{1+\beta}\left(d\Psi+\cos{\theta} d\phi\right)^{2}
    +d\Omega^{2}_{(2)}\right] \, .
\end{equation}

\noindent
Therefore, the horizon is a squashed 3-sphere and the squashing
parameter $\beta$ is given by Eq.~(\ref{eq:betadef}).

In the D1 solution, $\beta$ vanishes when both the total angular momentum $\mathcal J$
and the parameter ${\tilde\xi}_1$ vanish. Therefore, there can be squashing
even for vanishing total angular momentum due 
to the contribution of the dyonic field to the angular momentum at
the horizon (related to $\tilde{\xi }_{1}$).

In the D2 solution the squashing parameter can vanish even when
there is angular momentum
($\mathcal{J}_{\pm}=\mp\frac{\sqrt{3} \kappa^{2}\xi_{2}}{2g^{2}}$)
because there is a delocalized source of angular momentum in the dyonic
non-Abelian field.

\item The Bekenstein-Hawking entropy is given in terms of the
  near-horizon charges by

\begin{equation}
  S_{\rm{BH}}
  =
  \frac{A_{\rm H}}{4G_{N}^{(5)}}
  =
  \frac{\pi^{2}}{2 G_{N}^{(5)}}
  \sqrt{Q_{0} \tilde{\mathcal{Q}}_{+}\mathcal{Q}_{-}-4 \tilde{\mathcal{J} }^{2}} \, .
\end{equation}

Rewriting this expression in terms of the brane-source of Maxwell
charges is very difficult or would result in a very complicated
expression.

\item These black holes can be seen as non-Abelian generalizations of
  the 5-dimensional supersymmetric black holes of
  Ref.~\cite{Cvetic:1996xz} (with the  BMPV
   black hole
  \cite{Breckenridge:1996is} as a particular case). The non-Abelian
  interactions play an important role here, particularly in solution
  D2. They are the essential ingredient that allow us to describe an
  asymptotically flat, supersymmetric, rotating black hole with two
  different angular momenta, something that has not appeared so far in
  the literature. Furthermore, as a consequence of the interactions
  between electric and magnetic non-Abelian sources, the 2-form
  $d\omega$ is no longer anti-selfdual, as can be seen at the level of
  Eq.~(\ref{eq:omega}). This property, which does not hold here, was
  thought to be crucial to construct regular supersymmetric rotating
  black holes in five dimensions \cite{Herdeiro:2000ap}, although the
  analysis carried out in that reference did not include non-Abelian
  fields.

\item Even though the black holes are spinning, there is no
  ergosurface. This is expected for supersymmetric solutions because
  the existence of ergosurfaces was shown to be incompatible with
  supersymmetry in Ref.~\cite{Gauntlett:1998fz}.

\item The presence of closed timelike curves (CTCs) is a quite common
  feature of these kind of metrics. This problem has been studied with
  special emphasis in the context of the microstate geometries program
  \cite{Bena:2005va,Berglund:2005vb}. It turns out that the condition
  that guarantees the spacetime is free of closed timelike curves
  reduces to the positivity in the whole spacetime of certain
  function.\footnote{Recently in \cite{Avila:2017pwi},
  there has been some progress to reduce Eq.~(\ref{eq:absenceCTCs}) to
  an algebraic relation, simplifying the task of constructing explicit
  solutions. Although the results of \cite{Avila:2017pwi} only apply strictly to a special class of smooth horizonless solutions, we expect a similar analysis may also work for more general configurations.} In our case, we must demand

\begin{equation}\label{eq:absenceCTCs}
\mathcal{Z}_{0}\tilde{\mathcal{Z}}_{+}\mathcal{Z}_{-} H-\left(\omega_{5} H\right)^{2}- \left(\frac{{\breve \omega}_{\underline{\phi}}}{r\sin\theta}\right)^2 \ge 0\, .
\end{equation}

\noindent
In general, this is a complicated problem that has to be studied in a
case by case basis for particular values of the physical
constants. Often, however, it is enough to study this condition in the
$\rho\rightarrow 0$ and $\rho\rightarrow \infty$ limits, in which case it is
equivalent to the positivity of the horizon area $A_{\rm{H}}$ and
to the positivity of the ADM mass $\mathcal{M}$, respectively. In the case at hands, we have checked numerically that if this is the case, then Eq.~(\ref{eq:absenceCTCs}) can be satisfied without imposing more constraints on the parameters.

\item In the D1 solution, the instanton size $\kappa$ remains a modulus with
arbitrary value while the parameter $\tilde{\xi}_{1}$ appears, as we have seen, non-linearly in
the Maxwell and brane-source charges. It also contributes to
 some quantities computed at the horizon such as 
the angular momentum $\tilde {\mathcal J}$ and the entropy.

\item In the D2 solution, however,  the instanton size $\kappa$ is no longer a free parameter 
since it can be fixed for instance in terms of the electric charge of the dyon by using Eq. (\ref{eq:chargeD2}). 
As we have already seen, the non-Abelian fields of this solution also contribute to the 
total angular momentum in an asymmetric way, giving rise to different components
 of the angular momentum in different planes. Indeed, we can also use Eq. (\ref{eq:angularmomenta5d})
to fix the instanton size in terms of the combination of angular momenta $\Delta \mathcal J\equiv \mathcal J_+-\mathcal J_-$ as follows

\begin{equation}
\kappa^2=-\frac{g^2\Delta \mathcal J}{\sqrt{3}\xi_2}=-\frac{\sqrt{3}}{2 v}e^{\phi_\infty}k_{\infty}^{2/3} g \,\Delta \mathcal J \, .
\end{equation}

\end{itemize}
 
%%%%%%%%%%%%%%%%%%%%%%%%%%%%%%%%%%%%%%%%%%%%%%%%%%%%%%%%%%%%%%%%%%%%%%
%%%%%%%%%%%%%%%%%%%%%%%%%%%%%%%%%%%%%%%%%%%%%%%%%%%%%%%%%%%%%%%%%%%%%%
%%%%%%%%%%%%%%%%%%%%%%%%%%%%%%%%%%%%%%%%%%%%%%%%%%%%%%%%%%%%%%%%%%%%%%
\subsubsection*{Globally smooth solution}
%%%%%%%%%%%%%%%%%%%%%%%%%%%%%%%%%%%%%%%%%%%%%%%%%%%%%%%%%%%%%%%%%%%%%%
%%%%%%%%%%%%%%%%%%%%%%%%%%%%%%%%%%%%%%%%%%%%%%%%%%%%%%%%%%%%%%%%%%%%%%
%%%%%%%%%%%%%%%%%%%%%%%%%%%%%%%%%%%%%%%%%%%%%%%%%%%%%%%%%%%%%%%%%%%%%%

The family of solutions D2 includes a gobally regular and horizonless
solution that does not require the addition of localized brane sources
for the choice of charges
$\mathcal{Q}_{0}=\tilde{\mathcal{Q}}_{+}=\mathcal{Q}_{-}=\mathcal{J}=0$. In
this case, the $\mathcal{Z}$ functions now take the simpler form

\begin{eqnarray}
  \mathcal{Z}_{0}
  & = &
        1+\frac{2\, e^{-\phi_{\infty}} k_{\infty}^{2/3}}{3\, g^{2}}
        \frac{\rho^{2}+2\kappa^{2}}{\left(\rho^{2}+\kappa^{2}\right)^{2}}\, ,
  \\
\nonumber\\
  \mathcal{Z}_{-}
  & = &
        1\, ,
  \\
\nonumber\\
  \tilde{\mathcal{Z}}_{+}
  & = &
        1+\frac{9 \xi_{2}^{2}\, e^{\phi_{\infty}}k_{\infty}^{-2/3} }{ 2g^{2}}
        \frac{\left(2\rho^{2}+\kappa^{2}\right)\kappa^{2}}
        {\left(\rho^{2}+\kappa^{2}\right)^{2}}\, ,
\end{eqnarray}

\noindent
and the 1-form $\omega$ becomes just

\begin{equation}
  \omega
  =
 -\frac{\sqrt{3}\,  \xi_{2}\, \kappa^{2}\rho^{2}}{2g^{2}(\rho^{2}+\kappa^{2})^{2}}
      \left(d\phi+\cos{\theta} d\Psi\right)\, .
\end{equation}

This 5-dimensional solution describes the heterotic dyonic instanton
constructed in Ref.~\cite{Eyras:2000dg} compactified on a $T^{5}$. It
can also be seen as a rotating generalization of the instantonic
solution considered in Ref.~\cite{Cano:2017sqy}. The solution is
characterized by two non-vanishing asymptotic charges

\begin{eqnarray}
  \mathcal{Q}^{\infty}_{0}
  & = &
        \frac{2e^{-\phi_{\infty}}k_{\infty}^{2/3}}{3g^{2}}\, ,
  \\
  \nonumber\\
  \tilde{\mathcal{Q}}^{\infty}_{+}
  & = &
        \frac{9 \, e^{\phi_{\infty}}k_{\infty}^{-2/3}\, \xi_{2}^{2} \,\kappa^{2}}{g^{2}}\, ,
\end{eqnarray}

\noindent
and by only one independent angular momentum 

\begin{equation}
\mathcal {J}_{+}=-\mathcal {J}_{-}=-\frac{\sqrt{3} \,\xi_{2} \,\kappa^{2}}{2 \,g^{2}}\, .
\end{equation}

\noindent
Finally, the mass of the solution Eq.~(\ref{eq:mass5dbhs}) reduces to 

\begin{equation}
  \mathcal{M}
  =\, \frac{\pi}{4 G_N^{(5)}}
  \left(\mathcal {Q}_{0}^{\infty}+\mathcal {\tilde Q}_{+}^{\infty}\right)
  =
  \frac{\pi}{4 G_N^{(5)}}
  \left(\frac{2e^{-\phi_{\infty}}k_{\infty}^{2/3}}{3g^{2}}
    +\frac{9 \, e^{\phi_{\infty}}k_{\infty}^{-2/3}\, \xi_{2}^{2}\, \kappa^{2}}{g^{2}}\right)\, .
\end{equation}

%%%%%%%%%%%%%%%%%%%%%%%%%%%%%%%%%%%%%%%%%%%%%%%%%%%%%%%%%%%%%%%%%%%%%%
%%%%%%%%%%%%%%%%%%%%%%%%%%%%%%%%%%%%%%%%%%%%%%%%%%%%%%%%%%%%%%%%%%%%%%
%%%%%%%%%%%%%%%%%%%%%%%%%%%%%%%%%%%%%%%%%%%%%%%%%%%%%%%%%%%%%%%%%%%%%%
%%%%%%%%%%%%%%%%%%%%%%%%%%%%%%%%%%%%%%%%%%%%%%%%%%%%%%%%%%%%%%%%%%%%%%
\subsection{4-dimensional black holes $(a_{H}\neq0)$}
%%%%%%%%%%%%%%%%%%%%%%%%%%%%%%%%%%%%%%%%%%%%%%%%%%%%%%%%%%%%%%%%%%%%%%
%%%%%%%%%%%%%%%%%%%%%%%%%%%%%%%%%%%%%%%%%%%%%%%%%%%%%%%%%%%%%%%%%%%%%%
%%%%%%%%%%%%%%%%%%%%%%%%%%%%%%%%%%%%%%%%%%%%%%%%%%%%%%%%%%%%%%%%%%%%%%
%%%%%%%%%%%%%%%%%%%%%%%%%%%%%%%%%%%%%%%%%%%%%%%%%%%%%%%%%%%%%%%%%%%%%%

The 4-dimensional metric of these solutions is

\begin{equation}\label{eq:4drotatingbhs}
  ds_{(4)}^{2}
  =
  e^{2U}\left(dt+\breve{\omega}\right)^{2}
  -e^{-2U}\, \left(dr^{2}+r^{2}d\Omega^{2}_{(2)}\right)\, ,
\end{equation}

\noindent
where the 1-form $\breve{\omega}$ is given in Eq.~(\ref{eq:particularsolomega2}). The metric function
$e^{-2U}$ is

\begin{equation}
  e^{-2U}
  =
  \sqrt{\tfrac{27}{2} Z_{0}\tilde{Z}_{+} Z_{-} H-\left(\omega_{5} H\right)^{2}} \, ,
\end{equation}

\noindent
where

\begin{eqnarray}
  Z_{0}
  & = &
        a_{0}\left(1+\frac{q_{0}}{r}+\frac{2}{9 a_{0} a_{H}g^{2}}
        F\left(r;q_{H}, \lambda^{-2}\right)\right)\, ,
  \\
  \nonumber\\
  Z_{-}
  & = &
        a_{-}\left(1+\frac{q_{-}}{r}\right)\, ,
  \\
  \nonumber\\
  \tilde{Z}_{+}
  & = &
        \left\{\begin{array}{cc}
                 \displaystyle{
                 \tilde{a}_{+}\left(
                 1+\frac{{\tilde q}_{+}}{r}+\frac{4\xi_{1}^{2}}{\tilde{a}_{+}a_{-} g^{2}} \,
                 F\left(r;q_{-},\lambda^{-2}\right)
                 \right)
                 }\, ,
                 &
                   \text{(D1)}
                 \\
                 \\
                 \displaystyle{
                 \tilde{a}_{+}
                 \left(
                 1+\frac{{\tilde q}_{+}}{r}
                 +\frac{\xi_{2}^{2}}{\tilde{a}_{+}a_{-} g^{2}}
                 \frac{\left(r+q_{-}\right)
                 \left(1+2\lambda^{2} r\right)
                 +q_{-}\lambda^{4} r^{2}}{ \left (r+q_{-}\right)
                 \left(1+\lambda^{2} r\right)^{2}}
                 \right)
                 }\, ,
                 &
                   \text{(D2)}
\end{array}
                   \right.
  \\
  \nonumber\\
  H
  & = &
        a_{H}\left(1+\frac{q_{H}}{r}\right)\, ,
  \\
  \nonumber\\
  {\omega}_{5} & = &
                     \left\{
                     \begin{array}{cc}
                       \displaystyle{
                       a_{M}\left(
                       1+\frac{q_{M}}{r}
                       -\frac{2\sqrt{3}\,\xi_{1}}{a_{M} a_{H} g^{2}}\,
                       F\left(r; q_{H}, \lambda^{-2}\right)
                       \right)\, ,
                       }
                       &
                         \text{(D1)}
                       \\
                       \\
                       \displaystyle{
                       a_{M}\left(
                       1+\frac{q_{M}}{r}
                       -\frac{\sqrt{3}\,\xi_{2}}{2 a_{M} a_{H} g^{2}}
                       \frac{\lambda^{2}r\cos{\theta}}{\left(r+q_{H}\right)
                       \left(1+\lambda^{2}r\right)^{2}}
                       \right)\, ,
                       }
                       &
                         \text{(D2)}
                       \\
                     \end{array}
  \right.
\end{eqnarray}
where we have introduced the function

\begin{equation}
  F\left(r; q_{1}, q_{2}\right)
  \equiv
  \frac{\left(r+q_{1}\right)\left(r+2q_{2}\right)+q_{2}^{2}}
  {4q_{1} \left (r+q_{1}\right)\left(r+q_{2}\right)^{2}}\, .
\end{equation}

The relation between the constants $q_{0},{\tilde q}_{+}, q_{-}, q_{H}$ and
$q_{M}$ and the original parameters of the harmonic functions is

\begin{equation}
  q_{0}
  =
  \frac{1}{a_{0}}\left(b_{0}-\frac{1}{18b_{H}g^{2}}\right)\, ,
  \hspace{1cm}
  q_{-}
  =
  \frac{b_{-}}{a_{-}}\, ,
   \hspace{1cm}
  q_{H}
  =
  \frac{b_{H}}{a_{H}}\, ,
\end{equation}

\noindent
and 

\begin{equation}
  {\tilde q}_{+}
  =
  \left\{
    \begin{array}{cc}
      \displaystyle{
      \frac{1}{\tilde{a}_{+}}\left(b_{+}-\frac{\xi_{1}^{2}}{a_{-}g^{2}}\right)\, ,
      }
      &
        \text{(D1)}
      \\
      \\
      \displaystyle{
      \frac{b_{+}}{\tilde{a}_{+}}\, ,
      }
      &
        \text{(D2)}
      \\
    \end{array}
  \right. \, ,
\hspace{1.5cm}
q_{M}
=\left\{
  \begin{array}{cc}
    \displaystyle{
    \frac{1}{a_{M}}\left(b_{M}+\frac{\sqrt{3} \xi_{1}}{2g^{2} b_{H}}\right)\, ,
    }
    &
      \text{(D1)}
    \\
    \\
    \displaystyle{
    \frac{b_{M}}{a_{M}}\, ,
    }
    &
      \text{(D2)}
    \\
  \end{array}
\right. \, , 
\end{equation}

\noindent
where $\tilde{a}_{+}$ is again given by
Eq.~(\ref{eq:asymptoticvalueZ+}). We have implicitly assumed the
finiteness of several constants which appear in the denominators of
these expressions. In most cases, this is demanded by asymptotic
flatness, but we will have to take this fact into account at certain
points.

These 4-dimensional solutions depend on the parameters

\begin{equation}
  a_{0}, \tilde{a}_{+}, a_{-}, a_{H}, a_{M}, \lambda, g,
  q_{0}, \tilde{q}_{+}, q_{-}, q_{H}, q_{M},\,\,
  \text{and}\,\, \xi_{1}\,\, \text{or}\,\, \xi_{2}\, .
\end{equation}

\noindent
As already mentioned, not all of these parameters are independent
because they have to satisfy certain relations demanded by asymptotic
flatness and the standard normalization of the metric at spatial
infinity. These conditions are:

\begin{enumerate}
\item The vanishing of NUT charge. This condition demands that\footnote{This
    condition is also equivalent to imposing that the integrability condition
    of Eq.~(\ref{eq:breveomega}) is also satisfied at the pole. One may wonder if
    there is a fundamental reason to demand this since, after all, the $Z$ functions as well as the GH
    function $H$ are singular at that point. Leaving aside the requirement of
    asymptotic flatness and the wire singularities characteristic of Taub-NUT
    geometries \cite{Ortin:2015hya}, the main reason to impose the vanishing
    of the NUT charge is that we do not know of any string theory
    configuration (source) that can account for it. It was also argued in
    Ref.~\cite{Bellorin:2006xr} that the vanishing of the NUT charge is a
    necessary condition to for the solution to be globally supersymmetric.}

\begin{equation}
    \label{eq:ASFC14d}
a_{M} b_{H}=\,a_{H} \,b_{M}\, ,
\end{equation}

\noindent
which guarantees that $\breve{\omega}$ vanishes asymptotically, see
Eq.~(\ref{eq:particularsolomega2}). This equation can be satisfied in
two ways:

\begin{enumerate}

\item We can just set $a_{M} =b_{M}=0$.

\item If $a_{M}\neq 0$, we have to fix the integration constant
  $q_{M}$ in terms of $q_{H}$ (and $\xi_{1}$ in the D1 case) as
  follows

  \begin{equation}
    \label{eq:vanishingNUTcharge}
    q_{M}
    =
    \left\{\begin{array}{cc}
             \displaystyle{
             q_{H}+\frac{\sqrt{3}\xi_{1}}{2a_{M} a_{H} q_{H}g^{2}}\, ,
             }
             &
               \text{(D1)}
             \\
             \\
             q_{H}\, .
             &
               \text{(D2)}
           \end{array}
         \right.
\end{equation}

\end{enumerate}

Either way, when there is no NUT charge, the 1-form $\breve{\omega}$
is given by

\begin{equation}
  \label{eq:breveomegad4}
  \breve{\omega}
  =
  \left\{\begin{array}{cc}
           0\, ,
           &
             \text{(D1)}
           \\
\\
           \displaystyle{
           -\frac{\sqrt{3}\xi_{2}}{2g^{2}}
           \frac{\lambda^{2}r\sin^{2}\theta}{\left(1+\lambda^{2}r\right)^{2}}\,
           d\phi\, .
           }
           &
             \text{(D2)}
         \end{array}
       \right.
\end{equation}

Therefore, as already observed in Ref.~\cite{Meessen:2017rwm}, the
solution D1 is static, but the solution D2 describes a supersymmetric,
asymptotically-flat, rotating black hole.

\item At spatial infinity, the metric function $e^{2U}$ must take a
  constant value that is conventially taken to be $1$, \textit{i.e.},

\begin{equation}
  \lim_{r\rightarrow \infty} e^{2U}\rightarrow 1\, ,
  \qquad \Rightarrow\qquad
  \frac{27}{2}a_{0}\tilde{a}_{+}a_{-}a_{H}-\left(a_{M} a_{H}\right)^{2}=1\, ,
\end{equation}

\noindent
which allow us to rewrite $a_{0},\tilde a_{+}, a_{-}, a_{H}$ and
$a_{M}$ in terms of only four independent parameters. For future
convenience, we choose these parameters to be
$e^{\phi_{\infty}}, k_{\infty}, \ell_{\infty}$ and
$f_{\infty}$.\footnote{Recall that $\ell_{\infty}$ is the asymptotic
  value of the Kaluza-Klein scalar that measures the radius of the
  circle of the $5\rightarrow 4$ compactification. $f_{\infty}$ is the
  asymptotic value of the 5-dimensional metric function $f$, which is
  given in Eq.~(\ref{eq:fmetricfunction}) and no longer has to be
  equal to $1$.} The relation between these constants is

\begin{equation}
\begin{aligned}
  a_{0}
  & =
  \frac{1}{3}e^{\phi_{\infty}}k^{-2/3}_{\infty}f_{\infty}^{-1}\, ,
  \quad
  a_{-}=\,\frac{2}{3}e^{-\phi_{\infty}}
  k_{\infty}^{-2/3}f_{\infty}^{-1}\, ,
  \quad
  \tilde{a}_{+}=\,\,\frac{1}{3}k_{\infty}^{4/3}f_{\infty}^{-1}\, ,
  \\
  \\
  a_{H}
  & =
  \frac{f_{\infty}}{\ell_{\infty}}\, ,
  \quad
  a_{M}
  =\,\pm\frac{\ell_{\infty}}{f_{\infty}}
  \sqrt{\frac{1-\ell_{\infty} f_{\infty}^{2}}{\ell_{\infty} f_{\infty}^{2}}}\, ,
\end{aligned}
\end{equation}

\noindent
with $\ell_{\infty} f_{\infty}^{2}\le 1$. Together with the quotient $\xi_{2}/g$, these four
constants completely determine the asymptotic values of the
4-dimensional scalars which are given in
Eqs.~(\ref{eq:kkscalar5-4})-(\ref{eq:4dscalarA}). Defining

\begin{equation}
{Z}^{x}_{\infty}=v_{x}\,e^{i\gamma_{x}}\, , \qquad x=0,+,-,A\, ,
\end{equation}

\noindent
we find,

\begin{eqnarray}
  v_{0}
  & = &
        e^{-\phi_{\infty}}k_{\infty}^{2/3}\ell_{\infty}^{1/2}f_{\infty}^{-1}\, ,
        \qquad
        v_{+}
        =
        2k_{\infty}^{-4/3}\ell_{\infty}^{1/2}f_{\infty}^{-1}\, ,
  \\
\nonumber\\
  v_{-}
  & = &
        \left\{
        \begin{array}{lr}
          {\displaystyle
          e^{\phi_{\infty}}k_{\infty}^{2/3}\ell_{\infty}^{1/2}f_{\infty}^{-1}\, ,
          }
          &
            \hspace{.1cm}
            \text{(D1)}
          \\
          \\
          {\displaystyle
          e^{\phi_{\infty}}k_{\infty}^{2/3}\ell_{\infty}^{1/2}f_{\infty}^{-1}\,
          \left(
          1+\frac{9\, e^{\phi_{\infty}}f_{\infty}^{2} \,\xi_{2}^{2}}{2\, g^{2} \,k_{\infty}^{2/3}}
          \right)\, , 
          }
          &
            \hspace{.1cm}
            \text{(D2)}
        \end{array}
            \right.
  \\
  \nonumber\\ 
  v_{1}
  & = &
        v_{2}
        =
        0\, ,
        \qquad
        v_{3}
        =
        \left\{
        \begin{array}{lr}
          0\, ,
          &
            \hspace{.1cm}
            \text{(D1)}
          \\
          \\
          {\displaystyle
          \frac{2\, e^{\phi_{\infty}}\ell_{\infty}^{1/2}\,\xi_{2}}{g\, k_{\infty}^{2/3}f_{\infty}}\, , 
          }
          &
            \hspace{.1cm}
            \text{(D2)}
        \end{array}
            \right.
  \\
  \nonumber\\
  \tan^{2}\gamma_{x}
  & = &
        \frac{\ell_{\infty} f_{\infty}^{2}}{1-\ell_{\infty} f_{\infty}^{2}}\, ,
        \quad
        \forall x\, .
\end{eqnarray}
\end{enumerate}

Let us now rewrite the solution replacing the integration constants
$a_{0},a_{-},a_{M},\tilde{a}_{+},a_{H}$ by the physical parameters
$e^{\phi_{\infty}}, k_{\infty}, \ell_{\infty}$ and $f_{\infty}$. We
get\footnote{The $a_{M}=0$ case can be smoothly
  recovered in the $f_{\infty}\rightarrow 1/\sqrt{\ell_{\infty}}$
  limit. }

\begin{eqnarray}
  Z_{0}
  & = &
        \frac{e^{\phi_{\infty}} k_{\infty}^{-2/3}}{3\, f_{\infty}}
        \left(1+\frac{q_{0}}{r}+\frac{2\,e^{-\phi_{\infty}} {k}_{\infty}^{2/3}\ell_{\infty}}{3\,g^{2}}\,
        F\left(r;q_{H}, \lambda^{-2}\right)\right)\, ,
  \\
  \nonumber\\
  Z_{-}
  & = &
        \frac{2\,e^{-\phi_{\infty}} k_{\infty}^{-2/3}}{3\, f_{\infty}}
        \left(1+\frac{q_{-}}{r}\right)\, ,
  \\
  \nonumber\\
  \tilde{Z}_{+}
  & = &
        \left\{
        \begin{array}{lr}
          {\displaystyle
          \frac{k_{\infty}^{4/3}}{3\,f_{\infty}}
          \left(
          1+\frac{\tilde{q}_{+}}{r}
          +\frac{18\,e^{\phi_{\infty}}f_{\infty}^{2}\, \xi_{1}^{2}}{ \,k_{\infty}^{2/3}g^{2}} \,
          F\left(r;q_{-},\lambda^{-2}\right)
          \right)\, ,
          }
          &
            \hspace{.1cm}
            \text{(D1)}
          \\
          \\
          {\displaystyle
          \frac{k_{\infty}^{4/3}}{3\,f_{\infty}}\left(1+\frac{\tilde{q}_{+}}{r}
          +\frac{9\,e^{\phi_{\infty}}f_{\infty}^{2}\, \xi_{2}^{2}}{2 \,k_{\infty}^{2/3}g^{2}}
          \frac{\left(r+q_{-}\right)
          \left(1+2\lambda^{2} r\right)
          +q_{-}\lambda^{4} r^{2}}{ \left(r+q_{-}\right)
          \left(1+\lambda^{2} r\right)^{2}}\right)\, ,
          }
          &
            \hspace{.1cm}
            \text{(D2)}
        \end{array}
            \right.
  \\
  \nonumber\\
  H
  & = &
        \frac{f_{\infty}}{\ell_{\infty}}\left(1+\frac{q_{H}}{r}\right)\, ,
  \\
  \nonumber\\
  \omega_{5}
  & = &
        \left\{
        \begin{array}{lr}
          {\displaystyle
          \pm\frac{\ell_{\infty}}{f_{\infty}}
          \sqrt{\frac{1-\ell_{\infty} f_{\infty}^{2}}{\ell_{\infty} f_{\infty}^{2}}}
          \left(1+\frac{q_{H}}{r}\right)
          +\frac{\sqrt{3}\ell_{\infty}\xi_{1}}{2\,g^{2} \,f_{\infty}\, q_{H} r}
          -\frac{2\sqrt{3}\,\ell_{\infty}\,\xi_{1}}{g^{2}\, f_{\infty}}\,
          F\left(r; q_{H}, \lambda^{-2}\right)\, ,
          }
          &
            \hspace{.1cm}
            \text{(D1)}
          \\
          \\
          {\displaystyle
          \pm\frac{\ell_{\infty}}{f_{\infty}}
          \sqrt{\frac{1-\ell_{\infty} f_{\infty}^{2}}{\ell_{\infty} f_{\infty}^{2}}}
          \left(1+\frac{q_{H}}{r}\right)
          -\frac{\sqrt{3}\,\ell_{\infty}\,\xi_{2}}{2\, g^{2}\, f_{\infty}}
          \frac{\lambda^{2}r\cos{\theta}}{\left(r+q_{H}\right)
          \left(1+\lambda^{2}r\right)^{2}}\, .
          }
          &
            \hspace{.1cm}
            \text{(D2)}
        \end{array}
        \right.
\end{eqnarray}

At this point, the solutions depend on

\begin{equation}
e^{\phi_{\infty}}, k_{\infty}, \ell_{\infty}, f_{\infty}, g, q_{0},
\tilde{q}_{+}, q_{-}, q_{H}, \lambda, \xi_{1}, \text{or}\,\,
\xi_{2}\, .
\end{equation}

\noindent
The first 5 of these and $\xi_2$ are moduli (asymptotic values of scalar fields
and gauge coupling constant). The 4 $q$s will be interpreted as
near-horizon charges and we still have to find the physical meaning of
$\lambda$ and $\xi_{1}$. Let us study the charges of these
solutions.

%%%%%%%%%%%%%%%%%%%%%%%%%%%%%%%%%%%%%%%%%%%%%%%%%%%%%%%%%%%%%%%%%%%%%%
%%%%%%%%%%%%%%%%%%%%%%%%%%%%%%%%%%%%%%%%%%%%%%%%%%%%%%%%%%%%%%%%%%%%%%
%%%%%%%%%%%%%%%%%%%%%%%%%%%%%%%%%%%%%%%%%%%%%%%%%%%%%%%%%%%%%%%%%%%%%%
\subsubsection*{Charges of the solutions}
%%%%%%%%%%%%%%%%%%%%%%%%%%%%%%%%%%%%%%%%%%%%%%%%%%%%%%%%%%%%%%%%%%%%%%
%%%%%%%%%%%%%%%%%%%%%%%%%%%%%%%%%%%%%%%%%%%%%%%%%%%%%%%%%%%%%%%%%%%%%%
%%%%%%%%%%%%%%%%%%%%%%%%%%%%%%%%%%%%%%%%%%%%%%%%%%%%%%%%%%%%%%%%%%%%%%

The black hole solutions that we have constructed are electrically
charged with respect to the 4 non-trivial Abelian vectors
$A^{0}_{(4)},A^{1}_{(4)},A^{\pm}_{(4)}$ in
Eqs.~(\ref{eq:4d-AKK})-(\ref{eq:4d-Apm}) and magnetically charged only
with respect to the KK vector $A^{0}_{(4)}$. Therefore, these
dyonic black holes have 5 conserved (Abelian) charges: 4 electric and
1 magnetic. Not all these charges are independent, though, as a
consequence of the zero-NUT-charge condition, but this is something
that depends on the definition of charges being used. For the
near-horizon charges, Eq.~(\ref{eq:vanishingNUTcharge}) leaves only 4
independent charges in the D2 case, since $q_M=q_H$. In the D1 case,
however, the near-horizon charge associated to the function $\omega_5$ is
given by $q_H+  \sqrt{\frac{\ell_\infty f_\infty}{1-\ell_\infty f_\infty^2}}
\frac{\sqrt{3}\xi_1}{2q_Hg^2}$ and therefore it is not fixed since $\xi_1$ is 
a free parameter. This strongly suggests that the quantity 
$\xi_1/\left(q_H g^2\right)$  could be interpreted as the electric charge of 
the non-Abelian dyon.

We can now compute the asymptotic charges
$q_{0}^{\infty}, \tilde{q}^{\infty}_{+}, q^{\infty}_{-},
q^{\infty}_{H}, q_{M}^{\infty}$, defined as 

\begin{eqnarray}
  q^{\infty}_{0}
  & = &
        \lim_{r\to \infty}r \left(\frac{Z_{0}}{a_{0}}-1\right)
        =
        q_{0}+\frac{e^{-\phi_{\infty}}k_{\infty}^{2/3}\ell_{\infty}}{6\, g^{2} \,q_{H}}\, ,
  \\
  \nonumber \\
  q^{\infty}_{-}
  & = &
        \lim_{r\to \infty}r \left(\frac{Z_{-}}{a_{-}}-1\right)=q_{-}\, ,
  \\
  \nonumber\\
  {\tilde{q}}^{\infty}_{+}
  & = &
        \lim_{r\to \infty}r \left(\frac{{\tilde{Z}}_{+}}{{\tilde a}_{+}}-1\right)
        =
        \left\{
        \begin{array}{lr}
          {\displaystyle
          \tilde{q}_{+}
          +\frac{9\, e^{\phi_{\infty}}k_{\infty}^{-2/3}\,\xi_{1}^{2}}{2\, g^{2}\,
          f_{\infty}^{2}\,{q}_{-}}\, ,
          }
          &
            \hspace{.1cm}
            \text{(D1)}
          \\
          \\
          {\displaystyle
          \tilde{q}_{+}
          +\frac{9\, e^{\phi_{\infty}}k_{\infty}^{-2/3}\,\xi_{2}^{2}}{2\, g^{2}\, f_{\infty}^{2}}\left(q_{-}+2\lambda^{-2} \right)\,  ,
          }
          &
            \hspace{.1cm}
            \text{(D2)}
        \end{array}
            \right.
  \\
\nonumber\\
  q^{\infty}_{H}
  & = &
        \lim_{r\to \infty}r \left(\frac{H}{a_{H}}-1\right)
        =
        q_{H}\, ,
  \\
\nonumber\\
  q^{\infty}_{M}
  & = &
        \lim_{r\to \infty}r \left(\frac{\omega_{5}}{a_{M}}-1\right)
        =
        q^{\infty}_{H}
        =
        q_{H}\, .
\end{eqnarray}

\noindent
In both the D1 and D2 cases, we find only 4 independent charges as a
consequence of Eq.~(\ref{eq:vanishingNUTcharge}).

%%%%%%%%%%%%%%%%%%%%%%%%%%%%%%%%%%%%%%%%%%%%%%%%%%%%%%%%%%%%%%%%%%%%%%
%%%%%%%%%%%%%%%%%%%%%%%%%%%%%%%%%%%%%%%%%%%%%%%%%%%%%%%%%%%%%%%%%%%%%%
%%%%%%%%%%%%%%%%%%%%%%%%%%%%%%%%%%%%%%%%%%%%%%%%%%%%%%%%%%%%%%%%%%%%%%
\subsubsection*{Mass and angular momentum}
%%%%%%%%%%%%%%%%%%%%%%%%%%%%%%%%%%%%%%%%%%%%%%%%%%%%%%%%%%%%%%%%%%%%%%
%%%%%%%%%%%%%%%%%%%%%%%%%%%%%%%%%%%%%%%%%%%%%%%%%%%%%%%%%%%%%%%%%%%%%%
%%%%%%%%%%%%%%%%%%%%%%%%%%%%%%%%%%%%%%%%%%%%%%%%%%%%%%%%%%%%%%%%%%%%%%

Comparing the asymptotic expansion of the metric
Eq.~(\ref{eq:4drotatingbhs}) with

\begin{equation}
  ds^{2}_{(4)}
  \sim
  \left(1-\frac{2 G_{N}^{(4)} \mathcal{M}}{r}\right)dt^{2}
  +\frac{4\mathcal{J} \sin^{2} \theta}{r} dt d\phi
  -\left(1-\frac{2 G_{N}^{(4)} \mathcal{M}}{r}\right)dr^{2}
  -r^{2} d\Omega^{2}_{(2)}\, ,
\end{equation}

\noindent 
we find that the ADM mass $\mathcal{M}$ and the angular momentum
$\mathcal{J}$ of the solution are given by

\begin{eqnarray}
  \mathcal{M}
  & = &
        \frac{q_{0}^{\infty} +q_{+}^{\infty} +q_{-} +q_{H}
        -4\left(1-\ell_{\infty} f_{\infty}^{2}\right)\,q_{H}}{4 \,\ell_{\infty} f_{\infty}^{2}\,G_{N}^{(4)}}\, ,
  \\
  \nonumber \\
  \nonumber \\
  \mathcal J
  & = &
        \left\{\begin{array}{cc}
                 0\, ,
                 &
                   \hspace{2cm}\text{(D1)}
                 \\
                 \\
                 \displaystyle{-\frac{\sqrt{3}\,\xi_{2}}{4\,g^{2}\lambda^{2}}}\, .
                 &
                   \hspace{2cm} \text{(D2)}
                 \\
               \end{array}
  \right. 
\end{eqnarray}

\noindent
In the D2 case we can use this last equation to fix the instanton size $\lambda$ in
terms of angular momentum and the moduli of the solution:

\begin{equation}\label{eq:lambda}
\lambda^{-2} = -\frac{4g^{2}}{\sqrt{3} \xi_2} \mathcal J\, .\hspace{1.5cm} \text{(D2)}
\end{equation}

%%%%%%%%%%%%%%%%%%%%%%%%%%%%%%%%%%%%%%%%%%%%%%%%%%%%%%%%%%%%%%%%%%%%%%
%%%%%%%%%%%%%%%%%%%%%%%%%%%%%%%%%%%%%%%%%%%%%%%%%%%%%%%%%%%%%%%%%%%%%%
%%%%%%%%%%%%%%%%%%%%%%%%%%%%%%%%%%%%%%%%%%%%%%%%%%%%%%%%%%%%%%%%%%%%%%
\subsubsection*{Properties of the solutions}
%%%%%%%%%%%%%%%%%%%%%%%%%%%%%%%%%%%%%%%%%%%%%%%%%%%%%%%%%%%%%%%%%%%%%%
%%%%%%%%%%%%%%%%%%%%%%%%%%%%%%%%%%%%%%%%%%%%%%%%%%%%%%%%%%%%%%%%%%%%%%
%%%%%%%%%%%%%%%%%%%%%%%%%%%%%%%%%%%%%%%%%%%%%%%%%%%%%%%%%%%%%%%%%%%%%%

\begin{itemize}

\item The D1 solution is an asymptotically-flat, static solution
  characterized by the independent physical parameters

\begin{equation}
e^{\phi_{\infty}}, k_{\infty}, \ell_{\infty}, f_{\infty}, g, q_{0},
\tilde{q}_{+}, q_{-}, q_{H}, \xi_1,\lambda\, .
\end{equation}

The first 5 are moduli and the next 5 can be interpreted as near-horizon charges.
The parameter $\lambda$ characterizing the instanton size can
be interpreted as non-Abelian hair.

\item The D2 solution is an asymptotically-flat, rotating solution
  characterized by the independent physical parameters

\begin{equation}
e^{\phi_{\infty}}, k_{\infty}, \ell_{\infty}, f_{\infty}, g,\xi_2,  q_{0},
\tilde{q}_{+}, q_{-}, q_{H}, \mathcal{J}\, .
\end{equation}

The first 6 are the moduli of the solution and the next 4 are the near-horizon charges.
Finally,  $\mathcal{J}$ is the angular momentum. As already discussed, in this solution 
the instanton  size $\lambda$ gets fixed in terms of $\mathcal J$ and some of the moduli
by  Eq. (\ref{eq:lambda}).

\item Both solutions have a spherical horizon at $r=0$ with area 

\begin{equation}
  A_{\rm H}
  =
  4\pi \sqrt{q_{0}\,q_{+}\,q_{-}\,q_{H}-\left(q_{M} \,q_{H}\right)^{2}}\, ,
\end{equation}

\noindent
as long as this quantity is real and finite, \textit{i.e.}, if
$q_{0}\,q_{+}\,q_{-}\,q_{H}>\left(q_{M}\, q_{H}\right)^{2}$. 
The angular momentum of the D2 solution does not modify 
the shape of the horizon  because
$\lim_{r\to 0}\breve{\omega} \rightarrow 0$, contrary to what we found
in the 5-dimensional case.

\item If, on top of having a regular horizon, the condition

\begin{equation}
  \frac{27}{2}Z_{0} \tilde{Z}_{+} Z_{-} H>\left(\omega_{5} H\right)^{2}
  \quad \text{if} \quad r
  \ge
  0\, , 
\end{equation} 

\noindent
is satisfied everywhere\footnote{We have checked numerically that it can be satisfied.}, so that the metric function $e^{-2U}\neq 0$,
these solutions will describe, respectively, a static (D1) and a
rotating (D2) asymptotically-flat black hole.

\end{itemize}

%%%%%%%%%%%%%%%%%%%%%%%%%%%%%%%%%%%%%%%%%%%%%%%%%%%%%%%%%%%%%%%%%%%%%%
%%%%%%%%%%%%%%%%%%%%%%%%%%%%%%%%%%%%%%%%%%%%%%%%%%%%%%%%%%%%%%%%%%%%%%
%%%%%%%%%%%%%%%%%%%%%%%%%%%%%%%%%%%%%%%%%%%%%%%%%%%%%%%%%%%%%%%%%%%%%%
%%%%%%%%%%%%%%%%%%%%%%%%%%%%%%%%%%%%%%%%%%%%%%%%%%%%%%%%%%%%%%%%%%%%%%
\section{Discussion}
\label{sec-discussion}
%%%%%%%%%%%%%%%%%%%%%%%%%%%%%%%%%%%%%%%%%%%%%%%%%%%%%%%%%%%%%%%%%%%%%%
%%%%%%%%%%%%%%%%%%%%%%%%%%%%%%%%%%%%%%%%%%%%%%%%%%%%%%%%%%%%%%%%%%%%%%
%%%%%%%%%%%%%%%%%%%%%%%%%%%%%%%%%%%%%%%%%%%%%%%%%%%%%%%%%%%%%%%%%%%%%%
%%%%%%%%%%%%%%%%%%%%%%%%%%%%%%%%%%%%%%%%%%%%%%%%%%%%%%%%%%%%%%%%%%%%%%
\subsection{Summary of the results}

In this paper we have presented a general class of supersymmetric
solutions of 4- and 5-dimensional SU$(2)$-gauged supergravities with 8 supercharges (that is:
$\mathcal{N}=1,d=5$ and $\mathcal{N}=2,d=4$ supergravities) using the
techniques developed in Refs.~\cite{Bellorin:2006yr,Bellorin:2007yp,Meessen:2015enl}.

The novel aspect of our solutions is the addition of delocalized sources of 
charge through non-trivial dyonic Yang-Mills fields. As we have seen, they play a fundamental 
role in our analysis since the angular momentum of the solutions (specially in 4 dimensions)
is related to the electric-type charges introduced by these fields.

In 5 dimensions, we have found two non-Abelian generalizations of the well-known 
BMPV black hole. One of these (D2) has two independent angular 
momenta, which is a feature that has not been observed in the literature before for 
supersymmetric asymptotically-flat black holes.

We have also constructed 4-dimensional asymptotically-flat black holes
which can be seen as the non-Abelian counterparts of the heterotic black holes studied in Ref. \cite{Cvetic:1995uj}. 
One of these has in fact a non-vanishing angular momentum and it is 
regular,  being this the first example (up to our knowledge) of this type. Actually, a ``no-go'' theorem 
had been proven in Ref.~\cite{Bellorin:2006xr} in the context of ungauged
$\mathcal{N}=2,d=4$ supergravity. Indeed, these theories have Abelian
vector fields only, and, with a single center, dyonic fields do not
give rise to angular momentum and any other sources of angular
momentum give rise to singularities. The rotating 4d solution that
we have constructed overcomes these problems because of the delocalized
nature of the dyonic non-Abelian fields, which do give rise to angular
momentum.

 \subsection{Future directions}

 Since the theories that we have considered here can be obtained from toroidal
 compactification of the 10-dimensional Heterotic supergravity, these
 solutions can be easily uplifted to 10 dimensions as it has been recently
 done in Ref.~\cite{Cano:2017qrq}. Heterotic supergravity, however, does not
 capture the complete set of first-order $\alpha'$-corrections, since it is
 well-known that the effective action of the Heterotic Superstring
 \cite{Bergshoeff:1988nn} also contains terms which are quadratic in the
 curvature of the torsionful spin connection. In some cases, it can be argued
 that the corrections introduced by those terms are small enough to be
 ignored, but, as shown in Refs.~\cite{Cano:2018qev,Cano:2018brq}, sometimes
 it is possible to compute them exactly (to that order, at least) if the
 results of Ref.~\cite{Chimento:2018kop} can be applied to the particular
 supergravity solution under consideration.

 Thus, it is natural to consider the embedding in Heterotic Superstring theory
 of these solutions and the $\alpha'$ corrections of these solutions. Work in
 these directions is in progress \cite{kn:OR}.

%%%%%%%%%%%%%%%%%%%%%%%%%%%%%%%%%%%%%%%%%%%%%%%%%%%%%%%%%%%%%%%%%%%%%%
%%%%%%%%%%%%%%%%%%%%%%%%%%%%%%%%%%%%%%%%%%%%%%%%%%%%%%%%%%%%%%%%%%%%%%
%%%%%%%%%%%%%%%%%%%%%%%%%%%%%%%%%%%%%%%%%%%%%%%%%%%%%%%%%%%%%%%%%%%%%%
%%%%%%%%%%%%%%%%%%%%%%%%%%%%%%%%%%%%%%%%%%%%%%%%%%%%%%%%%%%%%%%%%%%%%%
\section*{Acknowledgments}
%%%%%%%%%%%%%%%%%%%%%%%%%%%%%%%%%%%%%%%%%%%%%%%%%%%%%%%%%%%%%%%%%%%%%%
%%%%%%%%%%%%%%%%%%%%%%%%%%%%%%%%%%%%%%%%%%%%%%%%%%%%%%%%%%%%%%%%%%%%%%
%%%%%%%%%%%%%%%%%%%%%%%%%%%%%%%%%%%%%%%%%%%%%%%%%%%%%%%%%%%%%%%%%%%%%%
%%%%%%%%%%%%%%%%%%%%%%%%%%%%%%%%%%%%%%%%%%%%%%%%%%%%%%%%%%%%%%%%%%%%%%

TO would like to thank David Mateos for useful conversations. Both authors are
pleased to thank Pablo A.~Cano, Samuele Chimento and especially Pedro F.~Ram\'{\i}rez for many
fruitful discussions. We also thank the referee for taking the time to review the manuscript and for helpful comments about it. This work has been supported in part by the
MINECO/FEDER, UE grant FPA2015-66793-P and by the Spanish Research Agency
(Agencia Estatal de Investigaci\'on) through the grant IFT Centro de
Excelencia Severo Ochoa SEV-2016-0597.  The work of AR was further supported
by a ``Centro de Excelencia Internacional UAM/CSIC'' FPI pre-doctoral grant. TO
wishes to thank M.M.~Fern\'andez for her permanent support.

%%%%%%%%%%%%%%%%%%%%%%%%%%%%%%%%%%%%%%%%%%%%%%%%%%%%%%%%%%%%%%%%%%%%%%
%%%%%%%%%%%%%%%%%%%%%%%%%%%%%%%%%%%%%%%%%%%%%%%%%%%%%%%%%%%%%%%%%%%%%%
%%%%%%%%%%%%%%%%%%%%%%%%%%%%%%%%%%%%%%%%%%%%%%%%%%%%%%%%%%%%%%%%%%%%%%
%%%%%%%%%%%%%%%%%%%%%%%%%%%%%%%%%%%%%%%%%%%%%%%%%%%%%%%%%%%%%%%%%%%%%%
%%%%%%%%%%%%%%%%%%%%%%%%%%%%%%%%%%%%%%%%%%%%%%%%%%%%%%%%%%%%%%%%%%%%%%
%%%%%%%%%%%%%%%%%%%%%%%%%%%%%%%%%%%%%%%%%%%%%%%%%%%%%%%%%%%%%%%%%%%%%%

\end{document}